\documentclass[sigconf,nonacm]{acmart}
\usepackage{color,soul}
\AtBeginDocument{%
  \providecommand\BibTeX{{%
    \normalfont B\kern-0.5em{\scshape i\kern-0.25em b}\kern-0.8em\TeX}}}

\newcommand{\upfg}{\texttt{UPFG} }
\newcommand{\ufsg}{\texttt{UFSG} }
\newcommand{\unfsg}{\texttt{UNFSG} }

\begin{document}

\title[Fear Speech in Indian WhatsApp Groups]{``Short is the Road that Leads from Fear to Hate'':\\ Fear Speech in Indian WhatsApp Groups}

\author{Punyajoy Saha}
\email{punyajoys@iitkgp.ac.in}
\affiliation{%
  \institution{Indian Institute of Technology}
  \city{Kharagpur}
  \state{West Bengal}
  \country{India}
}

\author{Binny Mathew}
\email{binnymathew@iitkgp.ac.in}
\affiliation{%
  \institution{Indian Institute of Technology}
  \city{Kharagpur}
  \state{West Bengal}
  \country{India}
}

\author{Kiran Garimella}
\email{garimell@mit.edu}
\affiliation{%
  \institution{MIT Institute of Data, Systems and Society}
  \city{Cambridge}
  \country{USA}
}

\author{Animesh Mukherjee}
\email{animeshm@cse.iitkgp.ac.in}
\affiliation{%
  \institution{Indian Institute of Technology}
  \city{Kharagpur}
  \state{West Bengal}
  \country{India}
}

\begin{abstract}
WhatsApp is the most popular messaging app in the world.
Due to its popularity, WhatsApp has become a powerful and cheap tool for political campaigning being widely used during the 2019 Indian general election, where it was used to connect to the voters on a large scale. Along with the campaigning, there have been reports that WhatsApp has also become a breeding ground for harmful speech against various protected groups and religious minorities. Many such messages attempt to instil fear among the population about a specific (minority) community. 
According to research on inter-group conflict, such `fear speech' messages could have a lasting impact and might lead to real offline violence. 
In this paper, we perform the first large scale study on fear speech across thousands of public WhatsApp groups discussing politics in India. 
We curate a new dataset and try to characterize fear speech from this dataset. 
We observe that users writing fear speech messages use various events and symbols to create the illusion of fear among the reader about a target community. 
We build models to classify fear speech and observe that current state-of-the-art NLP models do not perform well at this task. 
Fear speech messages tend to spread faster and could potentially go undetected by classifiers built to detect traditional toxic speech due to their low toxic nature.
Finally, using a novel methodology to target users with Facebook ads, we conduct a survey among the users of these WhatsApp groups to understand the types of users who consume and share fear speech.
We believe that this work opens up new research questions that are very different from tackling hate speech which the research community has been traditionally involved in. We have made our code and dataset public\footnote{\url{https://github.com/punyajoy/Fear-speech-analysis}} for other researchers.

\end{abstract}


\maketitle

\section{Introduction}

The past decade has witnessed a sharp rise in cases of violence toward various religious groups across the world. According to a 2018 Pew Research report\footnote{\url{https://www.pewresearch.org/fact-tank/2018/06/21/key-findings-on-the-global-rise-in-religious-restrictions/}}, most cases of violence are reported against Christians, Jews and Muslims. 
The shooting at Christchurch, Pittsburgh synagogue incident and the Rohingya genocide are a few prominent cases of religion-centric violence across the globe. 
In most of these cases of violence, the victims were religious minorities and social media played a role in radicalizing the perpetrators. 
According to a recent report by the US commission mandated to monitor religious freedom globally\footnote{\url{https://www.business-standard.com/article/pti-stories/uscirf-recommends-india-13-others-for-countries-of-particular-concern-tag-india-rejects-report-120042801712_1.html}}, India is one of the 14 countries where religious minorities are constantly under attack. 
In India, most of the religious conflicts are between Hindus and Muslims~\cite{84DeadIn79:online} who form 79\% and 13\% of the overall population, respectively. Recently, differences in opinions about the Citizenship Amendment Bill (CAB) have led to severe conflicts between the two communities in various parts of India.\footnote{\url{https://www.bbc.com/news/world-asia-india-50670393}}

There is not one clear answer to why such tensions have increased in the recent past, though many reports indicate the role of social media in facilitating them. 
Social media platforms like Facebook and WhatsApp provide a cheap tool to enable the quick spread of such content online.
For instance, reports have shown that some recent cases of religious violence in India were motivated by online rumours of cattle smuggling or beef consumption\footnote{Cows are considered sacred by Hindus in India and beef trade has always been a contentious issue between Hindus and Muslims~\cite{sarkar2016sacred}.} on social media platforms~\cite{Religiou26:online}. 
Similarly, messages inciting  violence against certain groups spread across WhatsApp during the Delhi riots in 2020~\cite{HowWhats65:online}.

However, due to the strict laws punishing hate speech in India\footnote{\url{https://en.wikipedia.org/wiki/Hate_speech_laws_in_India}}, many users refrain from a direct call for violence on social media, and instead prefer a subtle ways of inciting the readers against a particular community.
According to~\citet{buyse2014words}, this kind of speech is categorized as `fear speech', which is defined as ``an expression aimed at instilling (existential) fear of a target (ethnic or religious) group''.
In these types of messages, fear may be generated in various forms. These forms include but are not limited to

 \begin{itemize}
     \item Harmful things done by the target groups in the past or present (and the possibility of that happening again). 
     \item A particular tradition of the group which is portrayed in a harmful manner. 
     \item Speculation showing the target group will take over and dominate in the future.
 \end{itemize}
 
 In this paper, we identify and characterize the prevalence of fear speech from messages shared on thousands of public WhatsApp groups discussing politics in India.

\begin{table}%
\centering
\caption{Examples of \textit{fear speech} (FS),  \textit{hate speech} (HS), and \textit{non fear speech} (NFS). We show how the fear speech used elements from \textcolor{red}{history}, and contains \textcolor{blue}{misinformation} to vilify Muslims. At the end, they ask the readers, to take action by \textcolor{orange}{sharing the post}.}
\small
\begin{tabular}{p{7cm}l}
\textbf{Text (translated from Hindi)} & \textbf{Label} \\\hline
Leave chatting and read this post or else all your life will be left in chatting. \textcolor{red}{In 1378, a part was separated from India, became an Islamic nation - named Iran} \dots and now \textcolor{blue}{Uttar Pradesh, Assam and Kerala are on the verge of becoming an Islamic state} \dots People who do \textit{love jihad}
--- is a Muslim. Those who think of ruining the country ---  Every single one of them is a Muslim !!!! Everyone who does not share this message forward should be a Muslim. \textcolor{orange}{If you want to give muslims a good answer, please share!!} We will finally know how many Hindus are united today !! & FS \\\hline
That's why I hate Islam! See how these mullahs are celebrating. Seditious traitors!! & HS \\\hline
A child's message to the countrymen is that Modi ji has fooled the country in 2014, distracted the country from the issues of inflationary job development to Hindu-Muslim and patriotic issues. & NFS \\\hline

\end{tabular}
\label{tab:example_fs_nfs}
\end{table}

\noindent\textbf{Difference to hate speech}: At a first glance, fear speech might appear similar to the well known problem of hate speech. Table \ref{tab:example_fs_nfs} provides an example comparing fear speech with hate speech. First we show a typical example of fear speech against Muslims. We can see the use of various historical information and related misinformation in the text. These portray the Muslim community as a threat, thus creating a sense of fear in the minds of the reader. One interesting thing to notice, is that there are no toxic statements in this post as well as in most other fear speech posts in general. This is different from hate speech,
which usually contains derogatory keywords~\cite{elsherief2018hate}. 

In order to further highlight the difference, we use an India specific hate lexicon~\cite{bohra-etal-2018-dataset} and a state-of-the-art hate speech detection model~\cite{aluru2020deep} and evaluate the fear speech dataset that we have developed in this paper (Section~\ref{sec:dataset}). Both of them perform poorly, scoring 0.53 and 0.49 as macro F1 scores respectively. This can be partially attributed to the toxicity in these hate speech datasets and the lexicon/models trying to mostly bank their predictions on that. Finally, empirical analysis using a toxicity classifier from the Perspective API~\cite{perspective}, shows that the 
toxicity score of hate speech texts (0.57) is significantly ($p$-value $<0.001$) higher than for our text containing fear speech (0.48). 
While clear guidelines have been laid out for hate speech, fear speech is not moderated as of now. 

We particularly focus on understanding the dynamics of fear speech against Muslims in the public WhatsApp groups.
Encrypted platforms like WhatsApp, where, unlike open platforms like Facebook or Twitter, there is no content moderation, facilitate the spread and amplification of these messages.
The situation is particularly dangerous in large political groups, which are typically formed by like-minded individuals who offer no resistance or countering to fear speech messages.
Such groups have been used to target the Muslim community several times.\footnote{\url{https://www.huffingtonpost.in/entry/whatsapp-hate-muslims-delhi_in_5d43012ae4b0acb57fc8ed80}}

Following the data collection strategy from Garimella et. al~\shortcite{Garimella_2020}, we  collected the data from over 5,000 such groups, gathering more than 2 million posts. Using this data, we manually curated a dataset of 27k posts out of which $\sim8,000$ posts were fear speech and $\sim19,000$ were non fear speech.
Specifically, we make the following contributions:

\begin{itemize}

    \item For the first time, our work quantitatively identifies the prevalence and dynamics of fear speech at scale, on WhatsApp, which is the most popular social media in India.
    \item We do this by curating a large dataset of fear speech. The dataset consists of 7,845 fear speech and 19,107 non fear speech WhatsApp posts. 
    The dataset will be made public after the completion of the review process of the paper. %
    \item We develop models that can automatically identify messages containing fear speech.
    \item Finally, using a novel, privacy preserving approach, we perform an online survey using Facebook ads to understand the characteristics of WhatsApp users who share and consume fear speech.
\end{itemize}

Our study highlights several key findings about fear speech. We observe that the fear speech messages have different properties in terms of their spread (fear speech is more popular), and content (deliberately focusing on specific narratives to portray Muslims negatively).
Next, we focus on users posting fear speech messages and observe that they occupy central positions in the network, which enables them to disseminate their messages better. Using Facebook ads survey, we observe that users in fear speech groups are more likely to believe and share fear speech related statements and are more likely to take anti-Muslim stance on issues.
We further develop NLP models for automatic fear speech classification. We find that even the state-of-the-art NLP models are not effective in the classification task.

\section{Related Work}

Speech targeting minorities has been a subject of various studies in literature. In this section we first look into  previous research that deals with various types of speech and highlight how fear speech is different from them. 

\noindent\textbf{Hate speech}. Extreme content online is mostly studied under the hood of hate speech. Hate speech is broadly defined as a form of expression that ``\textit{attacks or diminishes, that incites violence or hate against groups, based on specific characteristics such as physical appearance, religion, descent, national or ethnic origin, sexual orientation, gender identity or other,and it can occur with different linguistic styles, even in subtle forms or when humour is used} \cite{fortuna2018survey}". Most of the research in this space has looked into building models for detection of hate speech~\cite{zhang2018detecting,aluru2020deep}, characterising it~\cite{mathew2019temporal},  and studying counter-measures for hate speech~\cite{chung2019conan,sinem2020generating}. 
However, most studies use their own definition of hate speech and the lack of a single definition makes it a difficult annotation task~\cite{ross2017measuring} and subject to abuse. For instance, there are multiple incidents where governments used vague definitions of hate speech to create laws against free speech to punish or silence journalists and protesters~\cite{doi:10.1080/23743670.2020.1729832,bjornskov2019social,bisri2020indonesian}. 
In recent years, there has been a push by researchers to look into specific definitions of hate speech  like dangerous speech and fear speech so that hate speech can be tackled at a more granular level \cite{gagliardone2019extreme}.

\noindent\textbf{Dangerous speech}.
One of the sub-fields, dangerous speech~\cite{benesch2012dangerous} is defined as an expression that \textit{``have  a  significant  probability  of catalyzing  or  amplifying violence  by  one  group  against  another,  given  the  circumstances  in  which  they  were  made  or disseminated''}. The main challenge about dangerous speech is that it is very difficult to assess whether a statement actually \textit{causes} violence or not. 
The authors provide various other factors like the speaker, the environment, etc. as essential features to identify dangerous speech. These features are largely anonymous in the online world. %

\noindent\textbf{Fear speech}.
In their paper \citet{benesch2012dangerous}, defined fear as one of the features of dangerous speech. 
\citet{klein2017fanaticism} et al. claim that a large amount of discussion on race on platforms like Twitter is actually inflicted with fear rather than hate speech in the form of content such as \#WhiteGenocide, \#Blackcrimes, \#AmericaUnderAttack.
A recent UNESCO report\footnote{\url{https://en.unesco.org/news/dangerous-speech-fuelled-fear-crises-can-be-countered-education}} even argues that fear speech can also facilitate other forms of harmful content. 
However, fear speech was formally defined by \citet{buyse2014words} et al as an expression that attempts to instill a sense of fear in the mind of the readers. Though it cannot be pinpointed if fear speech is the cause of the violence, it lowers the threshold to violence. 
Most of the work in this space is based in social theory and qualitatively looks at the role of fear as a technique used in expressing hatred towards a (minority) community. 
Our work on the other hand, is the first that looks at fear speech at scale quantitatively.

\noindent\textbf{Islamophobia}.
Another aspect relevant to our study is the study of Islamophobia on social media.
Dictionary definition of Islamophobia refers to \textit{fear of Muslim community}, but over time, studies on Islamophobia have also covered a broad range of factors including, hate, fear and threat against the Muslim community\footnote{\url{https://en.wikipedia.org/wiki/Islamophobia}}. There are various works studying the problem at scale, most of them covering the hate side of the domain~\cite{yasseri2019detecting}. The perspective of fear in Islamophobia is well-established but there is very less work studying this issue~\cite{gottschalk2008islamophobia}. One of the works have tried to establish the difference between direct hate speech and indirect fear speech against Muslims~\cite{fear_work} but it is mostly a limited case study. Our work can be considered studying the fear component of Islamophobia, but our framework for annotating and characterizing fear speech can be used to study fear speech targeted toward other religious or ethnic groups.

\noindent\textbf{WhatsApp}.
WhatsApp could be a good target for bad actors who want to spread hatred towards a certain community at scale. On platforms like Twitter and Facebook, the platforms can monitor content being posted and hence provide content moderation tools and  countering mechanisms in place like suspension of the user and blocking of the post for limiting the use of harmful/hateful language. 
WhatsApp, on the other hand, is an end-to-end encrypted platform, where the message can be seen only by the end users. 
This makes the spread of any form of harmful content in such platforms much easier. 
For this study, we focus on the public WhatsApp groups which discuss politics. These groups only make up a small fraction of the overall conversations on WhatsApp, which are private.
However, political groups on WhatsApp are extremely popular in countries like India and Brazil~\cite{lokniti2018} and have been used to target minorities in the past.
The Supreme Court of India has held WhatsApp admins liable for any offensive posts found in groups they manage\footnote{\url{https://www.thehindu.com/news/national/other-states/whatsapp-admins-are-liable-for-offensive-posts-circulated/article18185092.ece}}. 
Such strict laws could be a reason for the cautious nature of the users about spreading offensive and deliberately hateful posts in public groups and instead opt for subtle fear speech which is indirect.

\citet{garimella2018whatapp} performed one of the earliest work on WhatsApp, where they devised a methodology to extract data from public WhatsApp groups. Following a similar strategy, other works have studied political interaction of users in various countries like India~\cite{caetano2018analyzing,resende2019mis}, Brazil~\cite{caetano2018analyzing} and Spain~\cite{sampietro2019emoji}. The other part of research on WhatsApp studies the spread of misinformation~\cite{reis2020detecting,garimella2020images} in the platform.
While, misinformation is a nuanced problem, \citet{arun2019whatsapp} argues that content that has an intent to harm (for e.g., hate speech or fear speech) is more dangerous.
Currently, there is no work that studies hateful content both explicitly or implicitly on WhatsApp at scale.

\section{Dataset}\label{sec:dataset}
In this section, we detail the data collection and processing steps that we undertook. Our analysis relies on a large dataset obtained from monitoring public groups on WhatsApp.\footnote{Any group on WhatsApp which can be joined using a publicly available link is considered a public group.}

\subsection{Data collection}
In this paper, we use the data collected from public WhatsApp groups from India discussing politics which usually have a huge interplay with religion~\cite{Manufact87:online,HowtheBJ16:online}. %
In order to obtain a list of such public WhatsApp groups we resorted to lists publicized on well-known websites\footnote{For example, \url{https://whatsgrouplink.com/}} or social media platforms such as Facebook groups. 
Due to the popularity of WhatsApp in India, political parties massively create and advertise such groups to spread their party message. These groups typically contain activists and party supporters, and hence typically act as echo chambers of information. Surveys show that one in six users of WhatsApp in India are a member of one of such public groups~\cite{lokniti2018}.

We used the data collection tools developed by \citet{garimella2018whatapp} to gather the WhatsApp data. With help from journalists who cover politics, we curated lists of keywords related to politicians and political parties. Using this list we looked up public WhatsApp groups on Google, Facebook and Twitter using the query ``\url{chat.whatsapp.com} + \textit{query}'', where \textit{query} is the name of the politician or political party. The keyword lists cover all major political parties and politicians all across India in multiple languages.
Using this process, we joined and monitored over 5,000 political groups discussing politics.
From these groups, we obtained all the text messages, images, video and audio shared in the groups.
Our data collection spans for around 1 year, from August 2018 to August 2019. This period includes high profile events in India, including the national elections and a major terrorist attack on Indian soldiers. The raw dataset contained 2.7 million text messages.

\subsection{Pre-processing} 

The dataset contains over a dozen languages. To make it easier to annotate, we first filtered posts by language to only keep posts in Hindi and English, which cover over 70\% of our dataset. 
Next, we applied simple techniques to remove spam. We randomly sampled 100 messages and manually found that 29\% of them were spam. These spam messages include messages asking users to sign-up for reward points, offers, etc.,  phishing links, messages about pornography, and click-baits. To filter out these messages, we generated a set of high precision lexicon\footnote{The lexicon can be found here: \url{https://www.dropbox.com/s/yfodqudzpc7sp82/spam_filtered_keywords.txt?dl=0}} that can suitably remove such messages. Since the spam removal method is based on a lexicon, it is possible that some spam messages are missed. To cross-check the quality of the lexicon, after the cleaning, we randomly sampled 100 data points again from the spam-removed dataset and only found 3 spam messages. Detailed statistics about this spam filtered dataset are reported in Table~\ref{tab:total_data}.

\begin{table}[!htpb]
\centering
\caption{Characteristics of our dataset.}
\label{tab:total_data}
\small
\begin{tabular}{p{5cm}|l}
\hline
\textbf{Features}   & \textbf{Count} \\\hline
\#posts             & 1,426,482  \\ \hline
\#groups present    & 5,010     \\ \hline
\#users present     & 109,542  \\ \hline
average users per group   & 30   \\ \hline
average messages per group& 284    \\ \hline
average length of a message (in words)  & 89\\ \hline
\end{tabular}
\end{table}

\if{0}
\begin{figure}[!htpb]
	\centering
	\includegraphics[width=0.5\linewidth]{Figures/language_distribution.pdf}
	\caption{Percentage of posts corresponding to different languages in the dataset. We consider the top 10 regional languages in addition to English.}
	\label{fig:language_distribution}
\end{figure}
\fi

Since messages contain emojis, links, and unicode characters we had to devise a pre-processing method that is capable of handling such variety of cases. 
For our analysis, we not only use the text messages, but also the emoji and links. So, we develop a pre-processing method which can extract or remove the particular entities in the text messages as and when required. When doing text analysis we remove the emojis, stop words and URLs using simple regular expressions. Further, to tokenize the sentences we lowercase the English words in the message and use a multilingual tokenizer from CLTK~\cite{johnsonetal2014}, as a single message can contain words from multiple languages. For emoji and URL analysis we extract the particular entities using specific extractors\footnote{\url{https://github.com/lipoja/URLExtract}}\textsuperscript{, }\footnote{\url{https://github.com/carpedm20/emoji/}} for these entities.

\subsection{Generating lexicons}

We use lexicons to identify the targets in a message. Since, we are trying to identify fear speech against Muslims we build a lexicon related to Muslims. At first, we create a seed lexicon which has words denoting Muslims for e.g \textit{Muslims}, \textit{Musalman} (in Hindi). 
Next, we tokenize each post in the dataset with the method mentioned in the pre-processing section into a list of tokens. Since the words representing a particular entity may contains $n$-grams, we consider an $n$-gram generator --- \textit{Phraser}\footnote{\url{https://radimrehurek.com/gensim/models/phrases.html}} to convert the list of tokens (unigrams) to $n$-grams. We consider only those $n$-grams which have a minimum frequency of 15 in the dataset and restrict $n$ to 3. Thus, each sentence gets represented by a set of n-grams where $n$ can be 1, 2 or 3. The entire corpus in the form of tokenized sentences is used to train a word2vec model with default parameters\footnote{\url{https://radimrehurek.com/gensim/models/word2vec.html}}.
We bootstrap each of the seed lexicon using the word2vec model. For each word/phrase in a particular seed lexicon we generate 30 similar words/phrase based on the embeddings from the word2vec model. We manually select the entity specific words and add them to the seed lexicon. Next this modified seed lexicon is again considered as the seed lexicon, and we redo the former steps. This loop continues until we are unable to find any more keywords to add to the lexicon. This way we generate the lexicon for identifying messages related to Muslims. The lexicon thus obtained  can be found at this url\footnote{\url{https://www.dropbox.com/s/rremody6gglmyb6/muslim_keywords.txt?dl=0}}.

\section{Annotating messages for fear speech}

We filtered our dataset for messages containing the keywords from our lexicon and annotated them for fear speech. %
The annotation process enumerates the steps taken to sample and annotate fear speech against Muslims in our dataset.

\subsection{Annotation guidelines}

\if{0}
\binny{Papers with annotation guidelines}

\begin{itemize}
    \item The Gab Hate Corpus: A collection of 27k posts annotated for hate speech
    \item Guidelines and Framework for a Large Scale Arabic Diacritized Corpus
    \item Guidelines and Annotation Framework for Arabic Author Profiling 
    \item Large Scale Arabic Error Annotation: Guidelines and Framework
    \item Measuring the Reliability of Hate Speech Annotations:The Case of the European Refugee Crisis
    \item https://sharedtasksinthedh.github.io/2017/10/01/howto-annotation/
\end{itemize}
\fi

\begin{figure}[!ht]
    \centering
    \includegraphics[width=\linewidth]{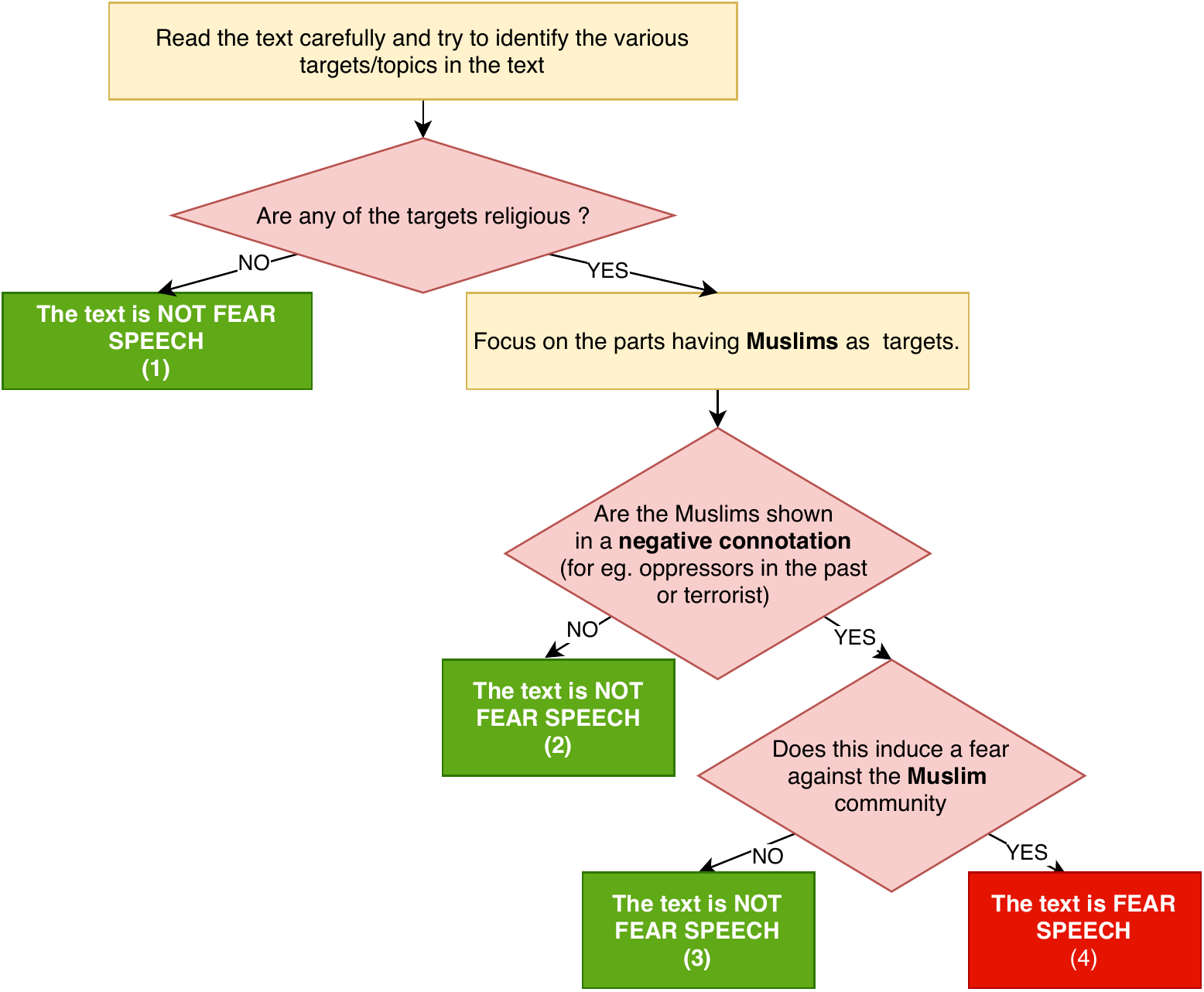}
    \caption{A step-by-step flowchart used by annotators to annotate any post as fear speech or non fear speech.}
    \label{fig:annotation}
\end{figure}

We follow the fear speech definition by Buyse et. al~\shortcite{buyse2014words} for our annotation process. In their work, fear speech is defined ``a\textit{s a form of expression aimed at instilling (existential) fear of a target (ethnic or religious) group}''. For our study, we considered the Muslims as the target group. In order to help and guide the annotators, we provide several examples highlighting different forms where they might find fear speech. These forms include but are not limited to (a) fear induced by using \textit{examples of past events}, e.g., demolition of a temple by a Mughal ruler, (b) fear induced by \textit{referring to present events}, e.g., Muslims increasing their population at an increasing rate. (c) fear induced by \textit{cultural references}, e.g., verses from the Quran, interpreted in a wrong way. (d) fear induced by \textit{speculation of dominance} by the target group, e.g., members of the Muslim community  occupying top positions in government institutions and  the exploitation of Hindus.\footnote{\url{https://thewire.in/communalism/sudarshan-news-tv-show-upsc-jihad-suresh-chavhanke-fact-check}} 
Figure~\ref{fig:annotation} shows the detailed flowchart used for the annotation process. 
Note that our annotation guidelines are strict and focused on a high precision annotation.
Further, we asked the annotators to annotate a post as fear speech even if only a part of the post appear to induce fear. This was done because many of the posts were long (as we see in Table~\ref{tab:total_data}, the average message has 89 words) and contained non fear speech aspects as well.

\subsection{Annotation training}

The annotation process was led by two PhD students as expert annotators and performed by seven under-graduate students who were novice annotators.
All the undergraduate students study computer science, and were voluntarily recruited through a departmental email list and compensated through an online gift card. 
Both the expert annotators had experience in working with harmful content in social media.

In order to train the annotators we needed a gold-label dataset. For this purpose the expert annotators annotated a set of 500 posts using the annotation guidelines. This set was selected by using the threshold of having at least one keyword from the Muslim lexicon. Later the expert annotators discussed the annotations and resolved the differences to create a gold set of 500 annotations. 
This initial set had 169 fear speech and 331 non fear speech post. From this set we sampled a random set of 80 posts initially for training the annotators. This set contained both the classes in equal numbers. After, the annotators finished this set of annotations we discussed  the incorrect annotations in their set with them.
This exercise further trained the annotators and fine-tuned the annotation guidelines. To check the effect of the first round of training, we sampled another set of 40 examples each from both classes again from the set of 500 samples. In the second round, most of the annotators could correctly annotate at least 70\% of the fear speech cases. The novice annotators were further explained about the mistakes in their annotations.

\subsection{Main annotation}
After the training process, we proceeded to the main annotation task by sampling posts having at least one keyword from the Muslim lexicon and gave them to the annotators in batches. For this annotation task we used the open source platform Docanno\footnote{\url{https://github.com/doccano/doccano}}, which was deployed on a Heroku instance. 
Each annotator was given a secure account where they could annotate and save their progress.

Each post was annotated by three independent annotators. They were instructed to read the full message and based on the guidelines provided, select the appropriate category (either fear speech or not). We initially started with smaller batches of 100 posts and later increased it to 500 posts as the annotators became well-versed with the task. We tried to maintain the annotators' agreement by sharing few of the errors in the previous batch. 
Since fear speech is highly polarizing and negative in nature the annotators were given ample time to do the annotations.

While there is no study which tries to determine the effect of fear speech annotations on the annotators, there are some evidence which suggest that the exposure to online abuse could lead to negative mental health issues~\cite{levin2017moderators,ybarra2006examining}.
Hence, the annotators were advised to take regular breaks and not do the annotations in one sitting. 
Finally, we also had regular meetings with them to ensure the annotations did not have any effect on their mental health.

\subsection{Final dataset}

Our final dataset consists of 4,782 posts with 1,142 unique messages labeled as fear speech and 3,640 unique messages labeled as not fear speech. 
We achieved an inter-annotator agreement of 0.36 using Fleiss $\kappa$ which is better than the agreement score on other related hate speech tasks~\cite{del2017hate,ousidhoum2019multilingual}. We assigned the final label using majority voting.

Next, we used locality sensitive hashing~\cite{gionis1999similarity} to find variants and other near duplicate messages of the annotated message in the dataset. 
Two documents were deemed to be similar if they have at least 7 hash-signatures matching out of 10.A group of similar messages following the former property will be referred to as shared message, henceforth. 
We manually verified 100 such messages and their duplicates and found error in 1\% of the cases.
This expanded our fear speech to $\sim8,000$ messages spread across $\sim1,000$ groups and spread by $\sim3,000$ users. 
Detailed statistics of our annotated dataset are shown in Table~\ref{tab:fear_speech_data}. %
Note that this dataset is quite different in its properties from the regular dataset shown in Table~\ref{tab:total_data}, with the average message being 5 times longer.

We observe that the non fear speech messages contain information pertaining to Quotes from Quran, political messages which talk about Muslims, Madrasa teachings and news, Muslim festivals  etc. An excerpt from one of the messages ``\textit{... Who is God?   One of the main beauties of Islam is that it acknowledges the complete perfection  greatness and uniqueness of God with absolutely no compromises....}''

\noindent\textbf{Ethics note}: We established strict ethics guidelines throughout the project. The Committee on the Use of Humans as Experimental Subjects at MIT approved the data collection as exempt.
All personally identifiable information was anonymized and stored separately from the message data.
Our data release conforms to the FAIR principles~\cite{wilkinson2016fair}.
We explicitly trained our annotators to be aware of the disturbing nature of social media messages and to take regular breaks from the annotation.

\begin{table}[!htpb]
\centering
\caption{Statistics of fear speech (FS) and non fear speech (NFS) in the annotated data.}
\label{tab:fear_speech_data}
\small
\begin{tabular}{p{5cm}|l|l}
\hline
\textbf{Features}   & \textbf{FS} & \textbf{NFS} \\\hline
\#posts             & 7,845  & 19,107 \\\hline
\#unique posts      & 1,142  &3,640 \\\hline
\#numbers of groups    & 917 & 1541 \\\hline
\#number of users     & 2,933 & 5,661 \\\hline
Average users per group    & 70 &  60 \\\hline
Average length of a message (in words)  & 500 & 464 \\\hline
\end{tabular}
\end{table}

\section{Analysis}
We use the dataset in Table~\ref{tab:fear_speech_data} to characterise fear speech at two levels: message level and user level. 
To further understand the user behaviour, we also conducted a survey among the WhatsApp group members to understand their perception and beliefs about fear speech. 
To measure statistical significance, we perform Mann-U-Whitney tests~\cite{mann1947test}, which is known to be stable across sample size differences.
Error bars in the plots represent 95\% confidence intervals. 

\subsection{Message characteristics}
In this section, we investigate the spread and dynamics of fear speech messages.

\begin{figure}[!ht]
    \centering
    \includegraphics[width=\linewidth]{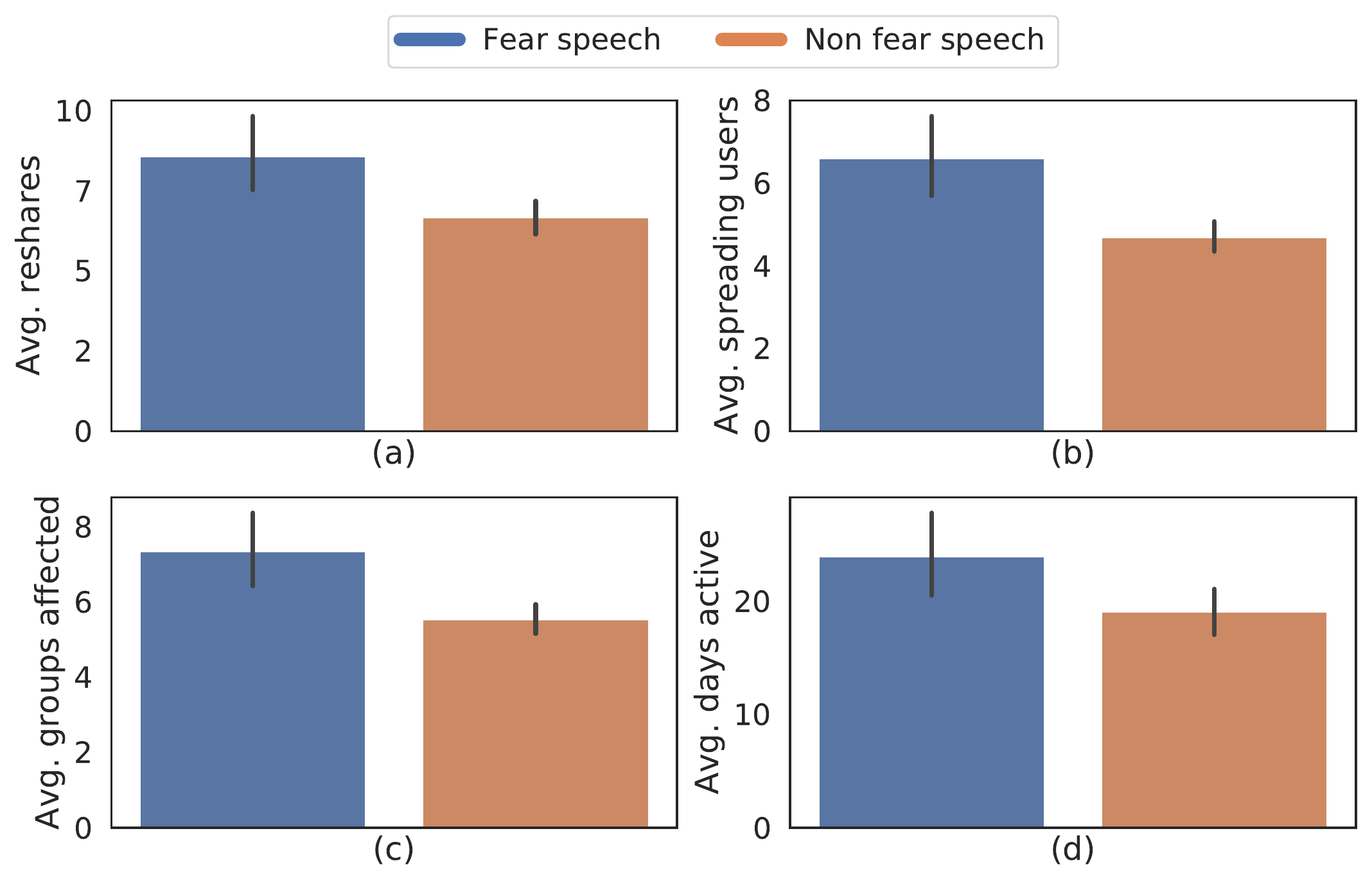}
    \caption{Characteristics of fear speech messages: (a) average number of times a message was shared $(p<0.001)$, (b) average number of users who shared a message $(p<0.0001)$,  (c) average number of groups the message was shared to $(p<0.0001)$, and (d) average number of days the message is active $(p<0.0001)$.}
    \label{fig:common_characteristics}
\end{figure}

\noindent\textbf{Spread characteristics}.
First, we compute the characteristics related to the spread of messages, like the number of shares. In Figure~\ref{fig:common_characteristics}(a), we observe that fear speech messages are shared more number of times on average as compared to non-fear speech messages. We also observe that these fear speech messages are spread by more number of users and sent to more groups on average (Figure~\ref{fig:common_characteristics}(b) \& Figure~\ref{fig:common_characteristics}(c), respectively).
Next, we compute the lifetime of a message as the time difference (in days) between the first and the last time the message was shared in our dataset. We observe that the lifetime of a fear speech message is more than that of a non fear speech message. All the differences shown in Figure \ref{fig:common_characteristics} are statistically significant. We also consider the fact that our dataset may be prone to right censoring i.e. the messages appearing close to the ending timestamp may reappear again. Hence, we consider all the messages appearing after June 2019 as message which can reappear a.k.a unobserved data. With the rest of the data, we trained survival functions~\cite{haslwanter2016analysis} for fear speech and non fear speech messages. Log rank test~\cite{mantel1966evaluation} on both the functions are significantly different $(p<0.0001)$.

These results suggest that fear speech, through its strong narratives and arguments, is able to bypass the social inhibitions of users and can spread further, faster and last longer on the social network.

\begin{figure}[!htpb]
    \centering
    \includegraphics[width=0.3\textwidth]{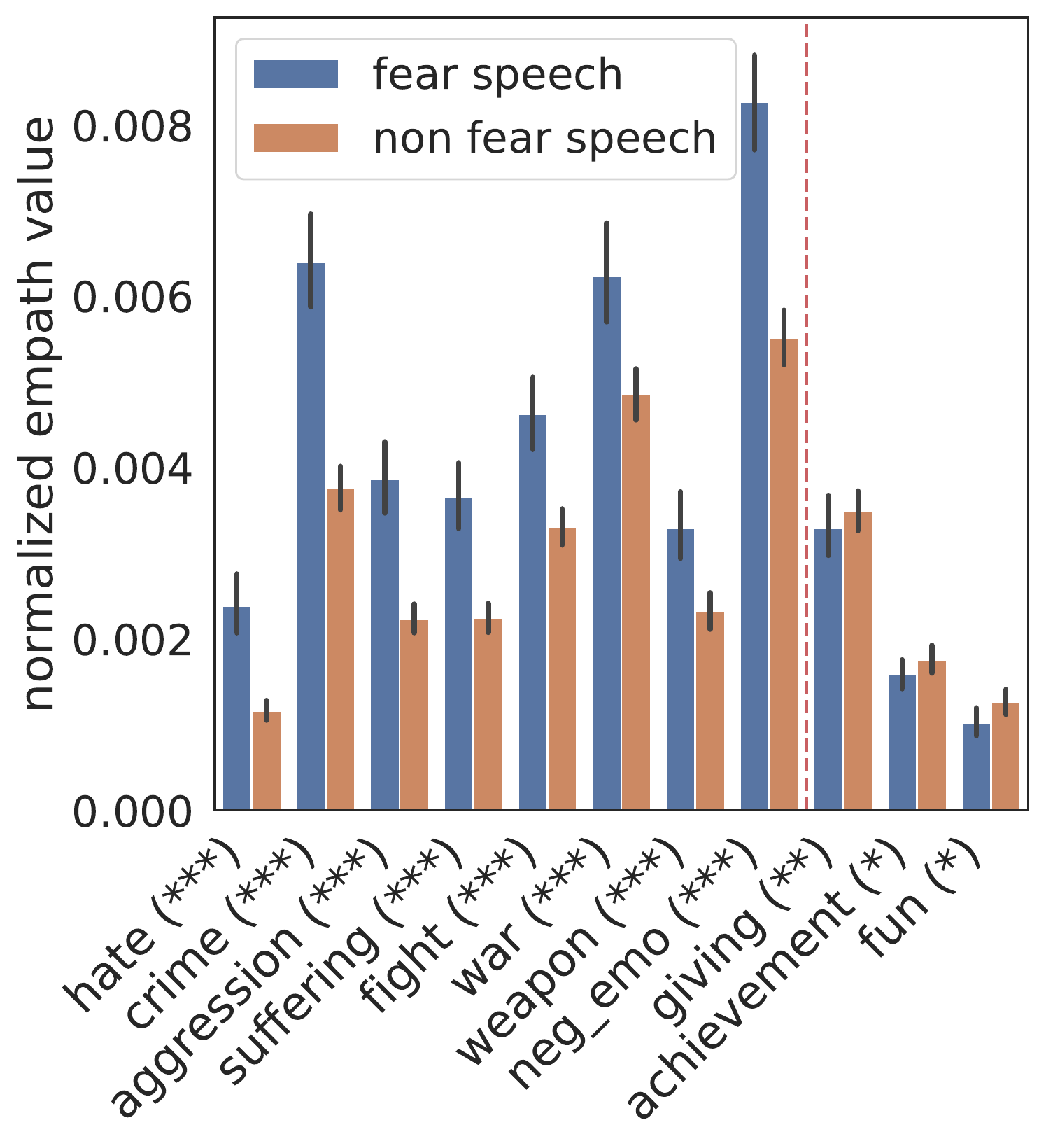}
    \caption{Lexical analysis using Empath. We report the mean values for several categories of Empath. Fear speech scored significantly high on topics like `hate', `crime', `aggression', `suffering', `fight', and `negative emotion (neg\_emo)' and `weapon'. Non fear speech scored higher on topics such as  `giving', `achievement' and `fun'. We use Mann-Whitney U \cite{mann1947test} test and show the significance levels ***$(p<0.0001)$,  **$(p<0.001)$, *$(p<0.01)$ for each of the category.}
    \label{fig:empath}
\end{figure}

\noindent\textbf{Empath analysis}.
Next, we perform lexical analysis using Empath~\cite{fast2016empath}, a tool that can be used to analyze text in over 189 pre-built lexical categories. First, we select 70 categories ignoring the topics irrelevant to fear speech for e.g. technology and entertainment. One this set of 70 categories, we characterize the english-translated version of the messages over these categories and report the top 10 significantly different categories in Figure~\ref{fig:empath}. Fear speech scores significantly high in topics like `hate', `crime', `aggression', `suffering', `fight', and `negative emotion (neg\_emo)' and 'weapon'. Non fear speech scores higher on topics such as  `giving', `achievement' and 'fun'. 
All the results are significant at least with $p$ \textless 0.01. To evaluate family-wise error rate we further applied sidak correction~\cite{seidler2000life} and observed all the categories except ``fun'' are still significant at 0.05.

\noindent\textbf{Topics in fear speech}.
To have a deeper understanding of the issues discussed in the fear speech messages, we use LDA \cite{hoffman2010online} to extract the topics as reported in Table~\ref{tab:topics_fear_speech}. Using the preprocessing methods explained earlier, we first clean numbers and URLs in the text. For each emoji, we add a space before and after it to separate the emojis which were joined. To get more meaningful topics, we use \textit{Phraser}\footnote{\url{https://radimrehurek.com/gensim/models/phrases.html}} to convert the list of tokens (unigrams) to bi-grams. We then pass the sentences (in the form of bi-grams) through the LDA model. To select the number of topics, we used the coherence score~\cite{10.1145/2684822.2685324} from 2 to 15 topics. We found that 10 topics received the highest coherence score of 0.45. Hence we used 10 topics for the LDA. Two of these topics were political and irrelevant to fear speech and hence ignored thus making 8 topics overall. 

Out of the topics selected, we clearly see a notion to promote negative thoughts toward the Muslim community portraying that they might be inciting disharmony across the nation. They discuss and spread various Islamophobic conspiracies around Muslims being responsible for communal violence (Topic $4$, $7$), to Muslim men promoting inter faith marriage to destroy the Hindu religion (Topic $5$). One of the topics also indicate exploitation of Dalits by the Muslim community (Topic $6$). In the annotated dataset, we found that Topic $6$ and $7$ were the most prevalent ones with 18\% of the posts belonging to each. The lowest number of posts (7\%) was found for Topic $5$.

\begin{table*}[!tbh]
\centering
\caption{List of topics discussed in fear speech messages.}
\small
\begin{tabular}{lp{12cm}p{3cm}}
 \textbf{Topic \#} & \textbf{Words} & \textbf{Name}\\
 \hline
1 & will kill, mosque, kill you, quran, \includegraphics[height=1em]{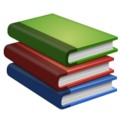}\includegraphics[height=1em]{Emoji/books_1f4da.png}, \includegraphics[height=1em]{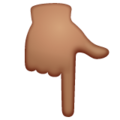}\includegraphics[height=1em]{Emoji/backhand-index-pointing-down_1f3fd.png}, love jihad, \includegraphics[height=1em]{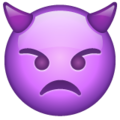}\includegraphics[height=1em]{Emoji/angry-face-with-horns_1f47f.png}, her colony, ramalingam, \includegraphics[height=1em]{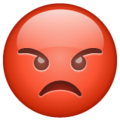}\includegraphics[height=1em]{Emoji/pouting-face_1f621.png}, \includegraphics[height=1em]{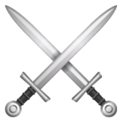}, crore, cattle, if girl, cleric, bomb bang, \includegraphics[height=1em]{Emoji/backhand-index-pointing-down_1f3fd.png}, children, war & Women mistreated in Islam \\\hline
2 & group, hindu brother, bengal, \includegraphics[height=1em]{Emoji/backhand-index-pointing-down_1f3fd.png}\includegraphics[height=1em]{Emoji/backhand-index-pointing-down_1f3fd.png}, temple, terrorist, rape, killing, zakat foundation, peoples, book hadith, page quran, vote, police, quran, against, indian, dalits, khwaja, story & UPSC jihad (Zakat foundation) \\\hline
3 & akbar, police, the population, \includegraphics[height=1em]{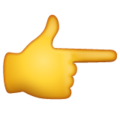}, islamic, society, war, gone, rape, population, children, family, love jihad, type, islamic, become & Muslim population \\\hline
4 & sri lanka, abraham, congress, love jihad, daughter, league, grandfather grandmother, university, family, girl, between, jats, i am scared, love, all, children, fear, pakistani, terrorist, \includegraphics[height=1em]{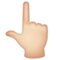}\includegraphics[height=1em]{Emoji/backhand-index-pointing-up_1f3fb.png}, without, \includegraphics[height=1em]{Emoji/backhand-index-pointing-right_1f449.png}& Sri Lanka Riots \\\hline
5 & temple smashed, answer, rape, \includegraphics[height=1em]{Emoji/backhand-index-pointing-right_1f449.png}, questions, gone, \includegraphics[height=1em]{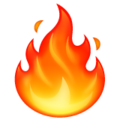}\includegraphics[height=1em]{Emoji/fire_1f525.png}, wrote, clean, girls, modi, woman, book hadith, hindu, whosoever won, will give, work, \includegraphics[height=1em]{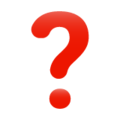}, \includegraphics[height=1em]{Emoji/pouting-face_1f621.png}\includegraphics[height=1em]{Emoji/pouting-face_1f621.png}, robbery& Love jihad \\\hline
6 &  congress, \includegraphics[height=1em]{Emoji/backhand-index-pointing-right_1f449.png}, type, about, savarkar, vote, dalit, indian, will go, islamic, he came, somewhere, our, leader, will do, terrorist, war, born, person, against, effort&Muslim exploitation of dalit \\\hline
7 & village, temple, kerala, quran, stay, become, mewat, history, between, congress, quran sura, family, mopala, rape, christian, sheela, dalit, living, om sai, \includegraphics[height=1em]{Emoji/books_1f4da.png}\includegraphics[height=1em]{Emoji/books_1f4da.png}, love jihad, earth, come, start& Kerala riots \\\hline
8 & congress, marathas, girl, delhi, kill, asura import, jihadi, \includegraphics[height=1em]{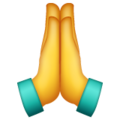}, master, janet levy, gives, mother father, surf excel, temple, \includegraphics[height=1em]{Emoji/pouting-face_1f621.png}\includegraphics[height=1em]{Emoji/pouting-face_1f621.png}, daughter, pigs, terrorists, maratha, century& Islamization of Bengal \\\hline
\end{tabular}
\label{tab:topics_fear_speech}
\end{table*}

\if{0}
\noindent\textbf{Temporal distribution of Fear speech} 

\binny{I am not observing anything interesting here. Should we include it? We can just keep it for now and later see if space allows. If you think there are interesting observations, we can add it}
\begin{figure}[!htpb]
    \centering
    \includegraphics[width=\linewidth]{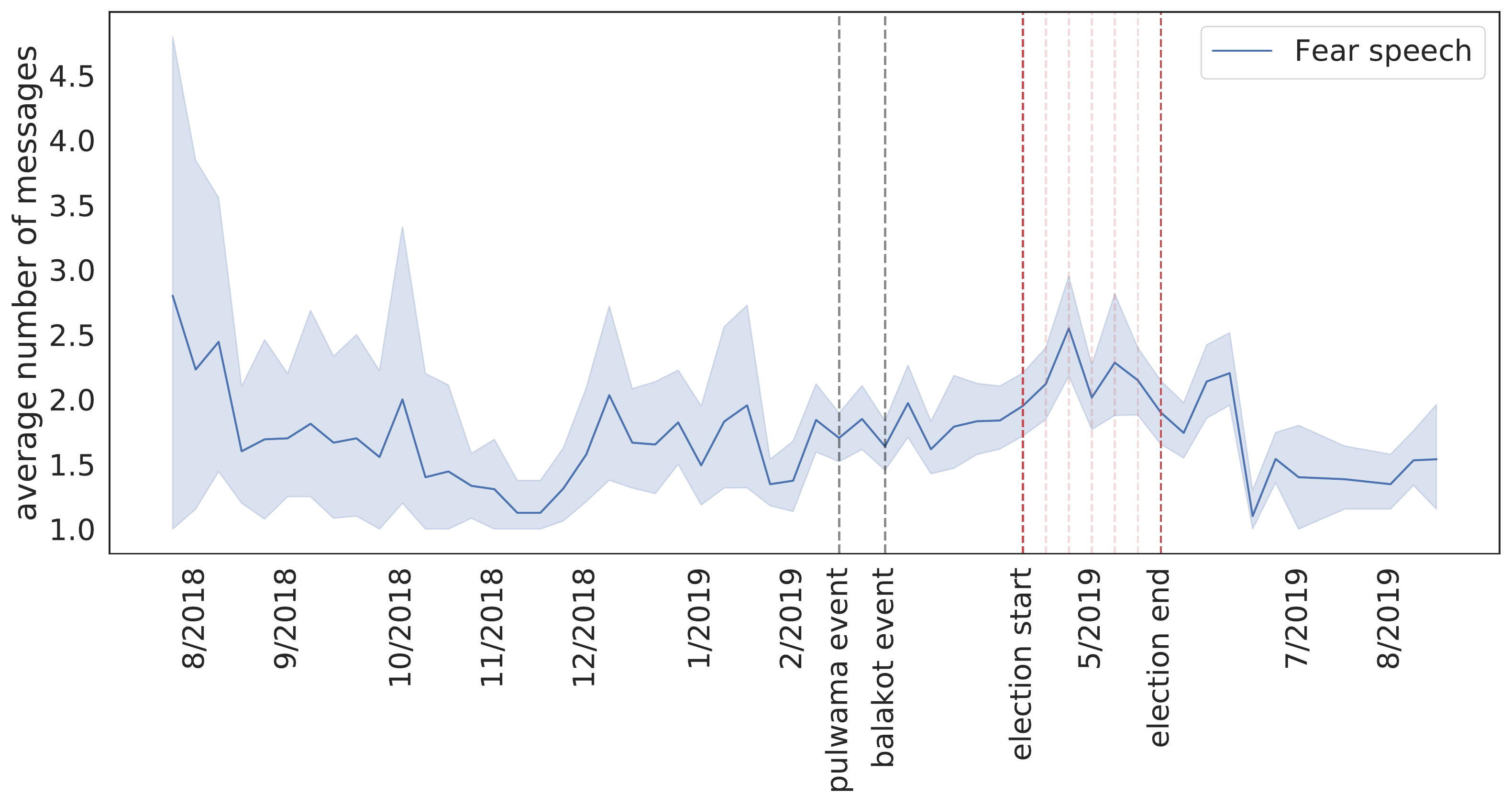}
    \caption{Temporal distrbution of fear speech}
    \label{fig:temporal distribution}
\end{figure}
\fi

\noindent\textbf{Emoji usage}.
We observe that 52\% of the fear speech messages had at least one emoji present in them, compared to 44\% messages if we consider the whole data.
Our initial analysis revealed that emojis were used to represent certain aspects of the narrative. For example, \includegraphics[height=0.9em]{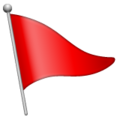} was used to represent the Hindutva (\textit{bhagwa}) flag\footnote{\url{https://en.wikipedia.org/wiki/Bhagwa_Dhwaj}}, \includegraphics[height=0.9em]{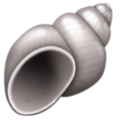} to represent purity in Hinduism\footnote{\url{https://www.boldsky.com/yoga-spirituality/faith-mysticism/2014/significance-of-conch-shell-in-hinduism-039699.html}}, \includegraphics[height=0.9em]{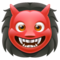} were used to demonize Muslims and  \includegraphics[height=0.9em]{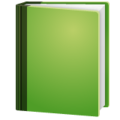} were used to represent the holy book of Islam, the Quran.
Further, these emojis also tend to frequently occur in groups/clusters. In order to understand their usage patterns we cluster the emojis. 
We first form the co-occurrence network~\cite{li2018co} of emojis where the nodes are individual emojis and edges represent that they co-occur within a window of $5$ characters at least once. The weight (W) of the edge is given by the equation \ref{eq:1}, where $F_{ij}$ represents the number of times the emojis $i$ and $j$ co-occur within a window of $5$ and $F_i$ represents the number of times emoji $i$ occurs.

\begin{equation}\label{eq:1}
    \text{W}_{ij}= F_{ij} / (F_i * F_j)
\end{equation}

After constructing this emoji network, we used the Louvain algorithm~\cite{blondel2008fast} to find communities in this network. 
We found 10 communities out of which we report the four most relevant communities in Table~\ref{tab:emoji_communities}.

We manually analyzed the co-occurrence patterns of these emojis found several interesting observations. Emojis such as \includegraphics[height=1em]{Emoji/triangular-flag_1f6a9.png}, \includegraphics[height=1em]{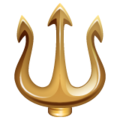}, \includegraphics[height=1em]{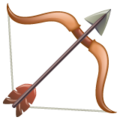},\includegraphics[height=1em]{Emoji/spiral-shell_1f41a.png}, \includegraphics[height=1em]{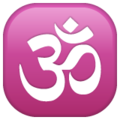},\includegraphics[height=1em]{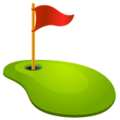} were used to represent the Hindutva ideology (row 1). Another set of emojis \includegraphics[height=1em]{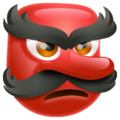}, \includegraphics[height=1em]{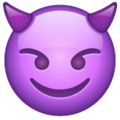}, \includegraphics[height=1em]{Emoji/angry-face-with-horns_1f47f.png}, \includegraphics[height=1em]{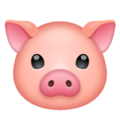}, \includegraphics[height=1em]{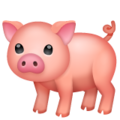} (row 2) was used to represent the Muslim community in a negative way. The former example helps in strengthening the intra-group (among members of the Hindu community) ties and the latter example vilifies the Muslim community as monsters or animals~\cite{buyse2014words}.

\begin{table}[!htpb]
\caption{\footnotesize{Top communities constructed using the Louvain algorithm from the emoji co-occurrence graph. Interpretation of the emojis as observed manually are added alongside each of the emoji communities.}}
 \label{tab:emoji_communities}
 \small
\begin{tabular}{p{5mm}p{4cm}p{3cm}}
 \hline Row& Emojis & Interpretation \\\hline
 1 & \includegraphics[height=1em]{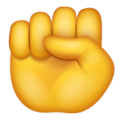},\includegraphics[height=1em]{Emoji/triangular-flag_1f6a9.png},\includegraphics[height=1em]{Emoji/folded-hands_1f64f.png},\includegraphics[height=1em]{Emoji/trident-emblem_1f531.png},\includegraphics[height=1em]{Emoji/bow-and-arrow_1f3f9.png},\includegraphics[height=1em]{Emoji/spiral-shell_1f41a.png},\includegraphics[height=1em]{Emoji/om_1f549.png},\includegraphics[height=1em]{Emoji/flag-in-hole_26f3.png},\includegraphics[height=1em]{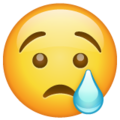}, \includegraphics[height=1em]{Emoji/backhand-index-pointing-up_1f3fb.png},\includegraphics[height=1em]{Emoji/fire_1f525.png},\includegraphics[height=1em]{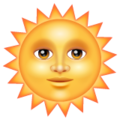},\includegraphics[height=1em]{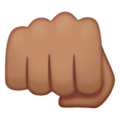},\includegraphics[height=1em]{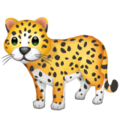},\includegraphics[height=1em]{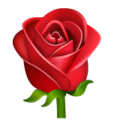},\includegraphics[height=1em]{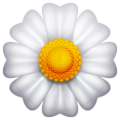},\includegraphics[height=1em]{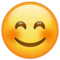},\includegraphics[height=1em]{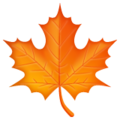}, \includegraphics[height=1em]{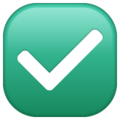},\includegraphics[height=1em]{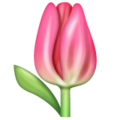},\includegraphics[height=1em]{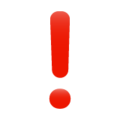},\includegraphics[height=1em]{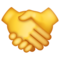},\includegraphics[height=1em]{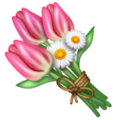},\includegraphics[height=1em]{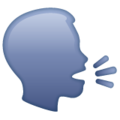},\includegraphics[height=1em]{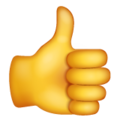},\includegraphics[height=1em]{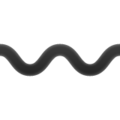},\includegraphics[height=1em]{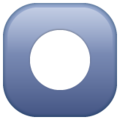} & Hindutva symbols\\\hline
 2& \includegraphics[height=1em]{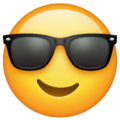},\includegraphics[height=1em]{Emoji/goblin_1f47a.png},\includegraphics[height=1em]{Emoji/green-book_1f4d7.png},\includegraphics[height=1em]{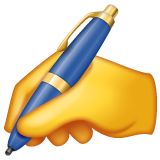},\includegraphics[height=1em]{Emoji/backhand-index-pointing-right_1f449.png},\includegraphics[height=1em]{Emoji/question-mark_2753.png},\includegraphics[height=1em]{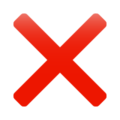},\includegraphics[height=1em]{Emoji/smiling-face-with-horns_1f608.png},\includegraphics[height=1em]{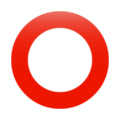}, \includegraphics[height=1em]{Emoji/angry-face-with-horns_1f47f.png},\includegraphics[height=1em]{Emoji/pig-face_1f437.png},\includegraphics[height=1em]{Emoji/pig_1f416.png},\includegraphics[height=1em]{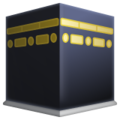},\includegraphics[height=1em]{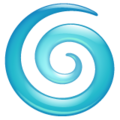},\includegraphics[height=1em]{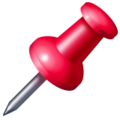},\includegraphics[height=1em]{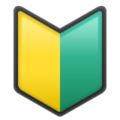} & Muslim as demons\\\hline
 3 & \includegraphics[height=1em]{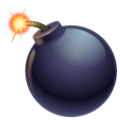},\includegraphics[height=1em]{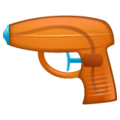},\includegraphics[height=1em]{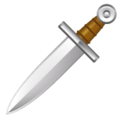},\includegraphics[height=1em]{Emoji/crossed-swords_2694.png},\includegraphics[height=1em]{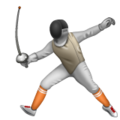},\includegraphics[height=1em]{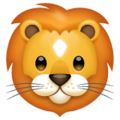},\includegraphics[height=1em]{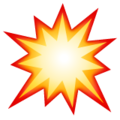},\includegraphics[height=1em]{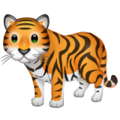},\includegraphics[height=1em]{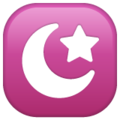}& terrorist attacks or riots by Muslims\\\hline
 4 & \includegraphics[height=1em]{Emoji/pouting-face_1f621.png},\includegraphics[height=1em]{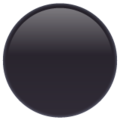},\includegraphics[height=1em]{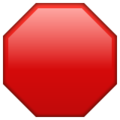},\includegraphics[height=1em]{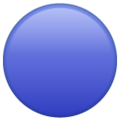},\includegraphics[height=1em]{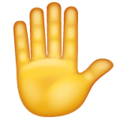},\includegraphics[height=1em]{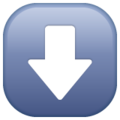}& Angry about torture on Hindus\\\hline
\end{tabular}
\vspace{-\baselineskip}
\end{table}

\begin{figure}[!ht]
    \centering
    \includegraphics[width=0.7\linewidth]{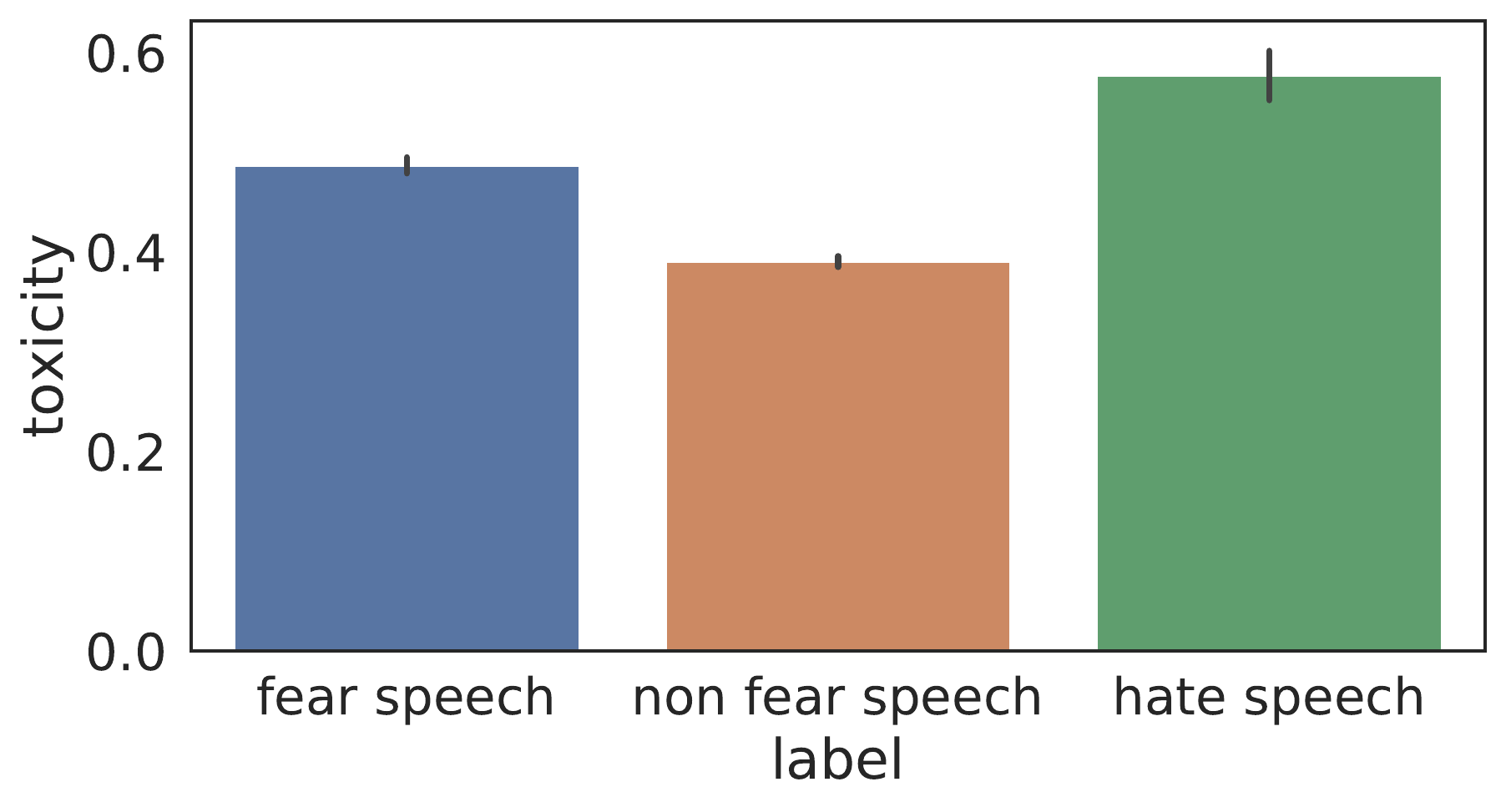}
    \caption{Toxicity comparison based on perspective api.}
    \label{fig:toxicity_comparison}
\vspace{-\baselineskip}
\end{figure}

\noindent\textbf{Toxicity}.
While qualitatively looking at the fear speech data, we observed that fear speech was usually less toxic in nature as compared to hate speech. 
To confirm this empirically, we used a recent hate speech dataset \cite{basile-etal-2019-semeval} and compared its toxicity with our dataset. Since targets of the posts were not annotated in the data, we used the English keywords in our Muslim lexicon to identify the hate speech targeting Muslims. Overall we found 155 hateful posts where one of the keywords from our lexicon matched. We passed the fear speech, non fear speech and hate speech subset dataset through the Perspective API~\cite{perspective}, which is a popular application for measuring toxicity in text. In Figure~\ref{fig:toxicity_comparison}, we observe that average toxicity of hate speech is higher than that of fear speech (p-value $<$ 0.001). Average toxicity of non fear speech is closer to fear speech. 
This shows how nuanced the problem at hand is and, thereby, substantiates the need for separate initiatives to study fear speech. 
In other words, while fear speech is dangerous for the society, the toxicity scores seem to suggest that the existing algorithms are not fine-tuned for their characterization/detection/mitigation.

To further establish the observation, we used a hate lexicon specific to Indian context~\cite{bohra-etal-2018-dataset} and measured its ability to detect fear speech. We assigned a \textit{fear speech} label for all the posts in our dataset where one or more keywords from the hate lexicon matched. Considering these labels as the predicted label, we got an F1 score of 0.53. Using a pre-trained hate speech detection model~\cite{aluru2020deep} and predicting the labels, also did not help as the pre-trained model performed more poorly (0.49). This clearly points out the need of novel mechanisms for the detection of fear speech.

\subsection{User characterization}
\begin{figure}[!htpb]
    \centering
    \includegraphics[width=0.7\linewidth]{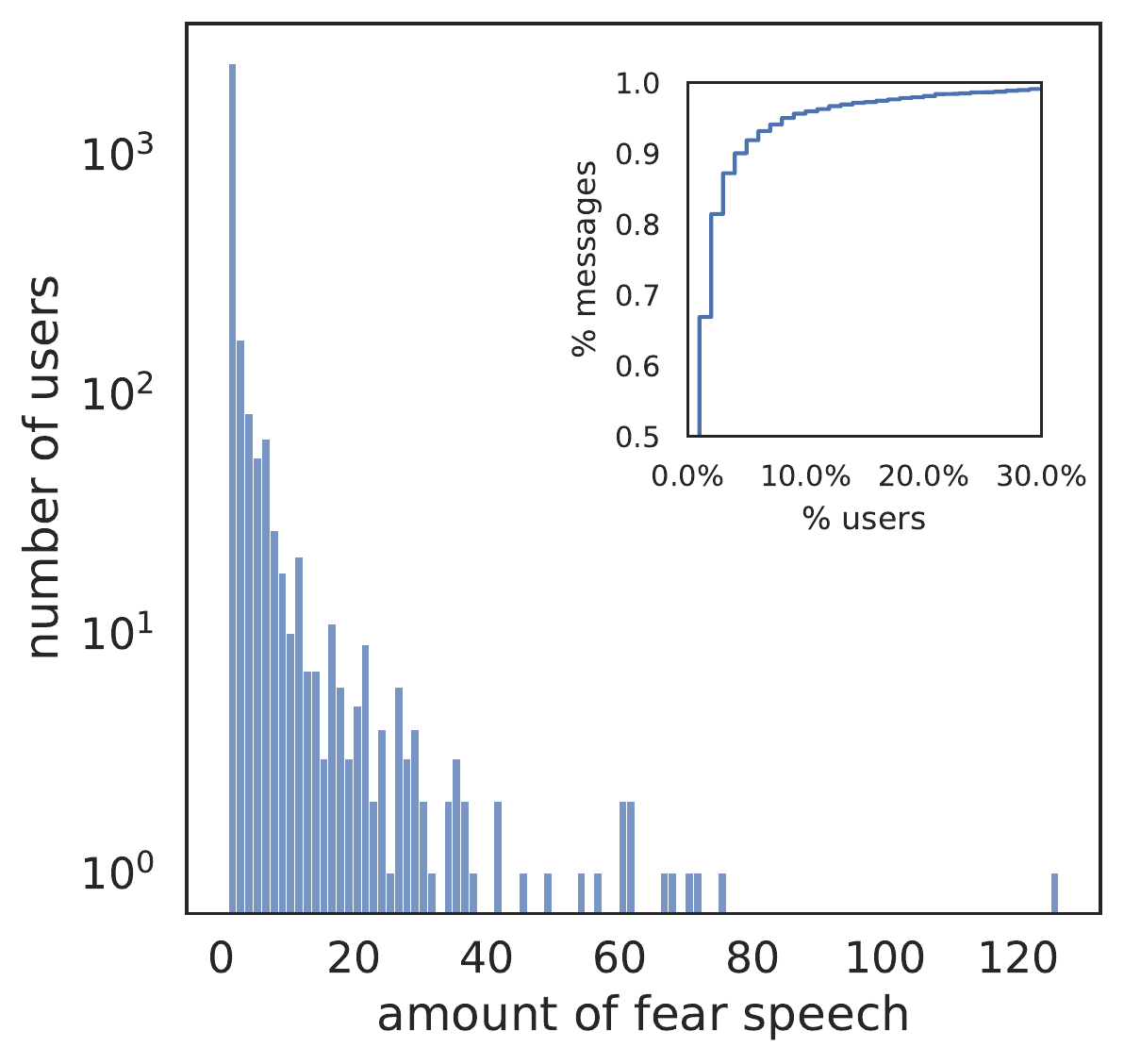}
    \caption{Distribution of fear speech messages posted by the fear speech users. Inset shows a CDF with cumulative \% of messages generated on the $y$-axis and user rank (converted to \%) on $x$-axis.}
    \label{fig:fearspeech_user_message}
\vspace{-\baselineskip}
\end{figure}

In this section, we focus on the $\sim3,000$ users who posted at least one of the fear speech messages to understand their characteristics. Figure~\ref{fig:fearspeech_user_message} shows the distribution of fear speech messages among the users. While most of these users post fear speech once or twice, there is a non-zero fraction of users who post 50+ times. Further only 10\% users posts around 90\% messages as shown in the inset of Figure~\ref{fig:fearspeech_user_message}. This indicates that there could be a hub of users dedicated for spawning such messages. We attempt to substantiate this through the core-periphery analysis below.

\noindent\textbf{Core-periphery analysis}. To understand the network positions of the fear speech users, we constructed a user-user network where there is link between two users if both of them are part of at least one group. The weight of the edge between two users represents the number of groups both of them are part of. This way we formed a network consisting of 109,292 nodes and 6,382,883 edges. 
To obtain a comparative set of users similar to the fear speech users, we sample a control set from the whole set of users (except the fear speech users) using propensity based matching~\cite{10.1093/biomet/70.1.41}. %
For matching we use the following set of features (a) avg. number of messages per month, (b) std. deviation of messages per month, (c) month the group had its first message after joining, and (d) the number of months the group had at least one message. We further measure the statistical significance between the fear speech users and the matched non fear speech users and found no significant difference ($p$ \textgreater 0.5).

\begin{figure}[!htpb]
    \centering
    \includegraphics[width=0.7\linewidth]{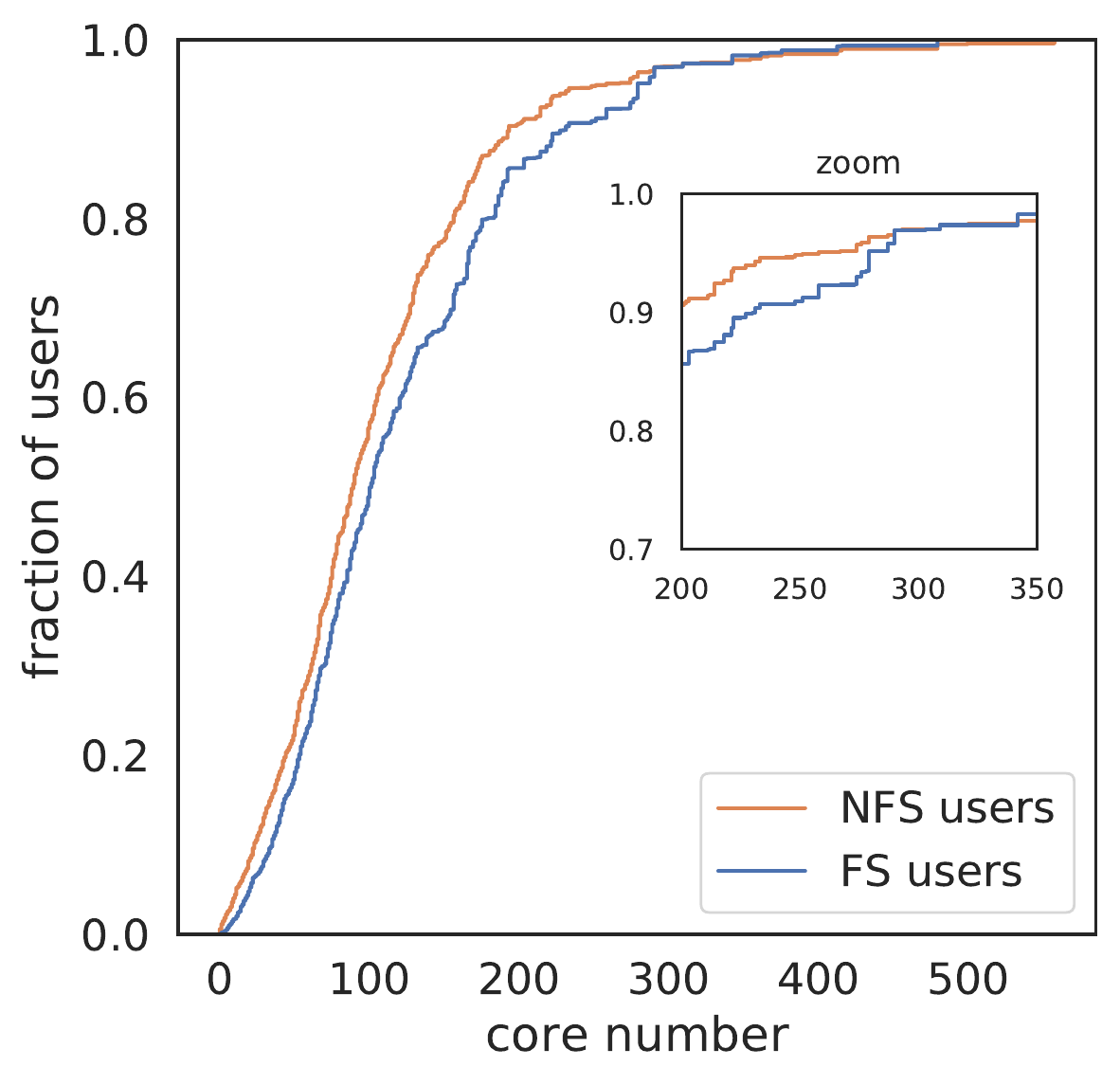}
    \caption{Cumulative distribution of $k$-core numbers for fear speech (FS users) and non fear speech users (NFS users). We see that FS users have a lower core number (higher centrality). The differences are significant at $(p<0.0001)$. A zoomed inset further highlights the difference between the two curves.}
    \label{fig:kcore_users}
\vspace{-\baselineskip}
\end{figure}

Next, we utilize $k$-core or coreness metric~\cite{shin2016corescope} to understand the network importance of the fear speech and non-fear speech users. Nodes with high coreness are embedded in major information pathways and have been shown to be influential spreaders, that can diffuse information to a large portion of the network~\cite{kitsak2010identification,malliaros2016locating}.
Figure~\ref{fig:kcore_users} shows the cumulative distribution of the core numbers
for the fear and the non fear speech users, respectively. We observe that the fear speech users are occupying far more central positions in the network, as compared to non fear speech users ($p$-value < 0.0001 with small effect size~\cite{effect_sizes} of 0.20). This indicates that some of the fear speech users constitute a hub-like structure in the core of the network. Further, 8\% of these fear speech users are also \textit{admins} in the groups where they post fear speech.

\if{0}\begin{figure}[!htpb]
    \centering
    \includegraphics[width=0.5\linewidth]{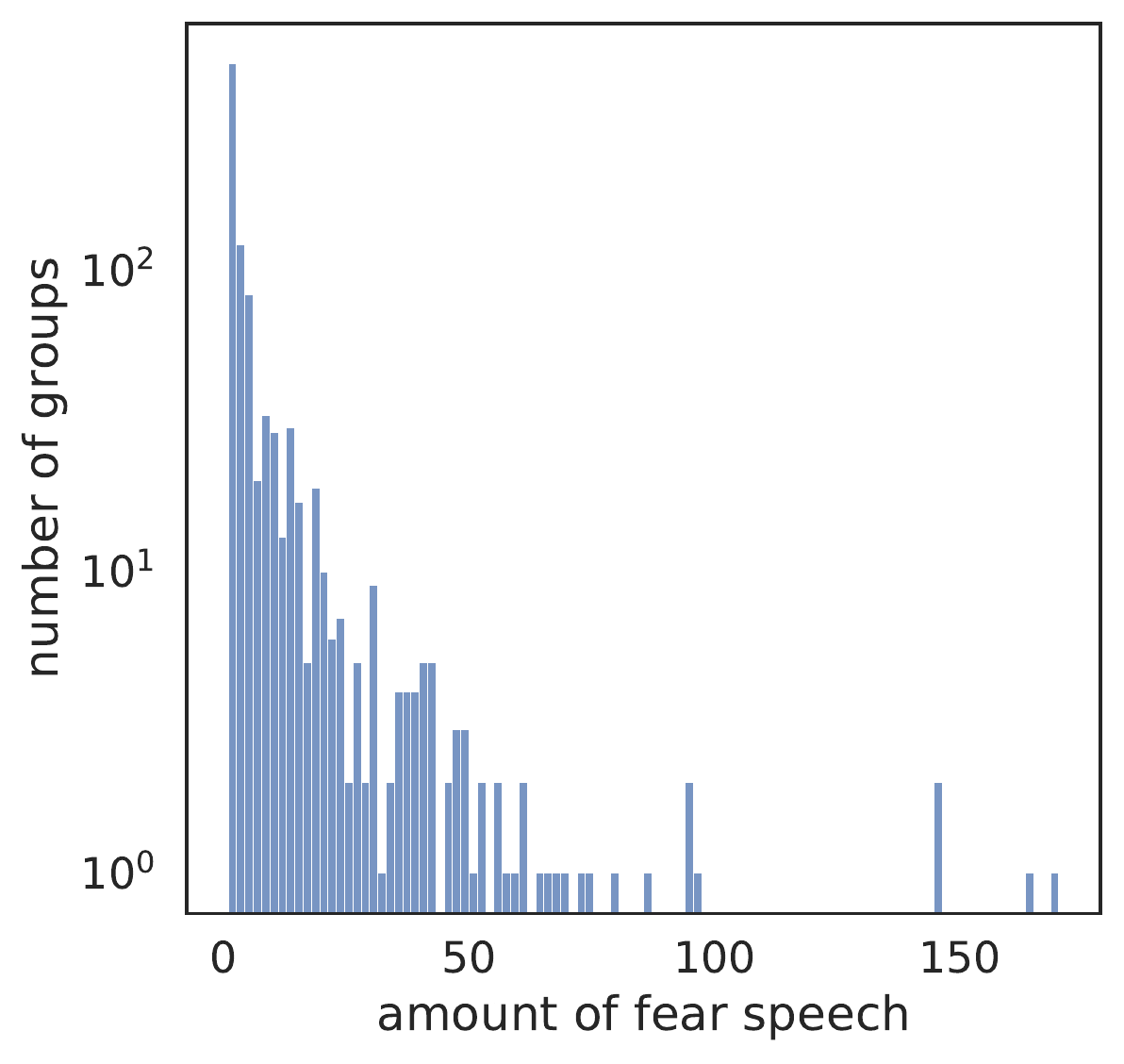}
    \caption{Distribution of fear speech messages among the fear speech groups.}
    \label{fig:fearspeech_group_message}
\end{figure}\fi

\subsection{Survey to characterize users}
In order to characterize users who share or consume fear speech, we used a novel, privacy preserving technique to survey our WhatsApp user set.
We used the Custom Audience targeting feature provided by Facebook\footnote{\url{https://www.facebook.com/business/help/341425252616329?id=2469097953376494}}, where the targeting is based on the lists of phone numbers that can be uploaded to Facebook.
Users having a Facebook account with a matching phone number can thus be targeted.\footnote{Note on privacy: The custom audience targeting only works for a matching user set of at least 1,000 phone numbers. So we cannot identify individual responses and tie them back to their WhatsApp data.}
To use this feature, we first created three sets of users:
(i) users who posted fear speech messages themselves (\upfg, around 3,000 users), (ii) users who were a part of one of the groups where fear speech was posted and did not post any fear speech message themselves (\ufsg, around 9,500 users from the top 100 groups posting the most fear speech), and, (iii) a controlled set of users who neither posted any fear speech nor were part of a group where it was posted in our dataset (\unfsg, around 10,000 users from a random sample of 100 groups which do not post any fear speech). Only roughly 50\% of these users had Facebook accounts with a matching phone number and were eligible for the advertisements to be shown.

\begin{figure}
    \centering
    \includegraphics[width=0.4\textwidth]{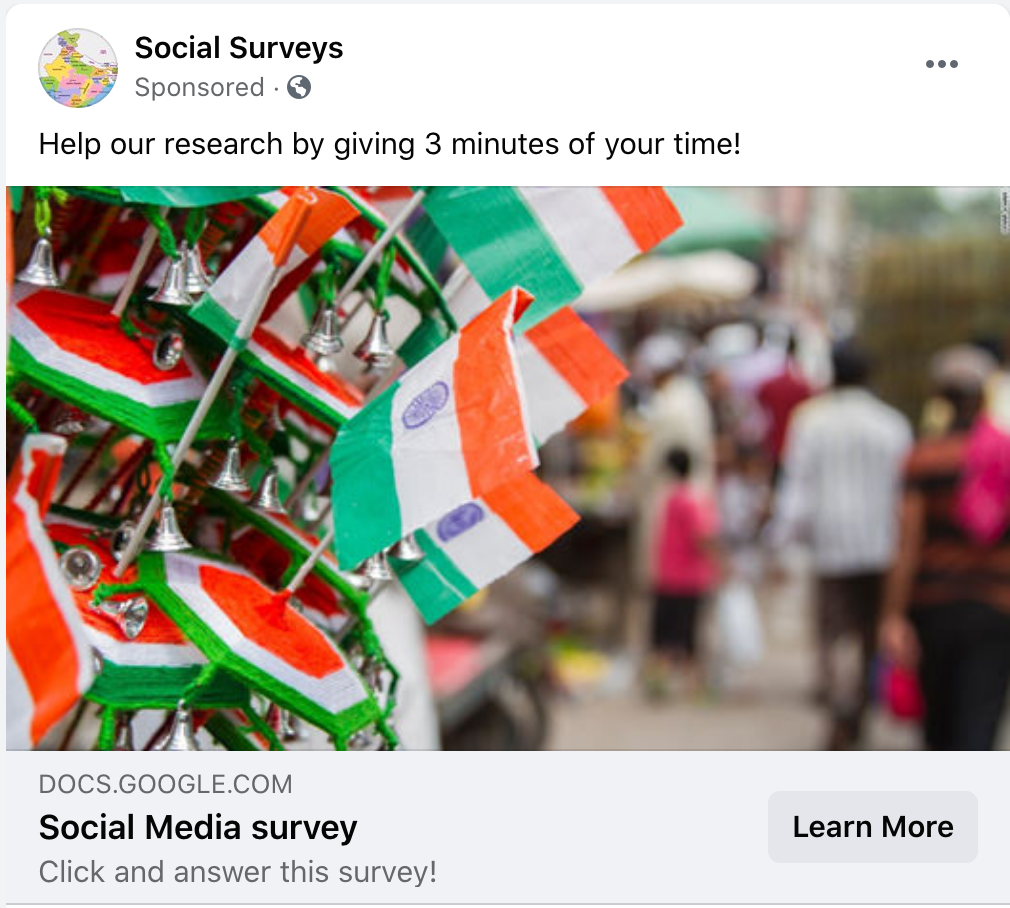}
    \caption{English translation of the ad used to obtain survey responses. All ads were run in Hindi.}
    \label{fig:survey_ad}
\vspace{-\baselineskip}
\end{figure}

We showed an ad containing a link to the survey (hosted on Google Forms). The survey was short and took at most 3 minutes of a user's time. No monetary benefits were offered for participating.
An English translation of the ad that was used is shown in Figure~\ref{fig:survey_ad}. 
To avoid any priming effects, we used a generic `Social media survey' prompt to guide the users to the survey. When the users would click the ad and reach our survey page, we first ask for consent and upon consent, the users are taken to the survey.

The survey was a 3x2 design: the three user sets presented with 2 types of statements: fear speech and non fear speech. \footnote{\url{https://www.dropbox.com/s/ajrrxt5k33mn3u3/survey_links.txt?dl=0}} 
In total, we chose 8 statements\footnote{\url{https://www.dropbox.com/s/y842qfnb81deo1q/survey_statements.txt?dl=0}} --- 4 containing carefully chosen snippets from fear speech text in our dataset and 4 statements containing true facts. To avoid showing overly hateful statements, we paraphrased the fear speech messages from our dataset to show a claim, e.g., \textit{`In 1761, Afghanistan got separated from India to become an Islamic nation'}.
The participants were asked whether they believed in the statement, and if they would share the statements on social media.
Along with the core questions, we had two optional questions about their gender and the political party they support.
Finally, to obtain a baseline on the beliefs of the various groups of users about Muslims, we asked 2 questions on their opinion about recent high profile issues involving Muslims in India. These include their opinion about (a) the Citizen Amendment Bill and (b) the Nizamuddin Markaz becoming a COVID-hotspot~\cite{HowTabli17:online}.

All the participants were shown fact checks containing links debunking the fear speech statements at the end of the survey.
To keep the survey short per user, we split the 8 statements into two surveys with four statements per survey with two being from the fear speech set and the other two being from the non fear speech set of our dataset.

The ads ran for just under a week, and we received responses from 119 users. 
The low response rate (around 1\%) is expected and was observed in other studies using Facebook ads to survey users~\cite{hoffman2020facebook} without incentives. 
A majority of the respondents (85\%) were male. Among the rest 5\% were female and 10\% did not disclose their gender.

\begin{figure}[!htpb]
    \begin{minipage}{.155\textwidth}
        \includegraphics[width=\linewidth]{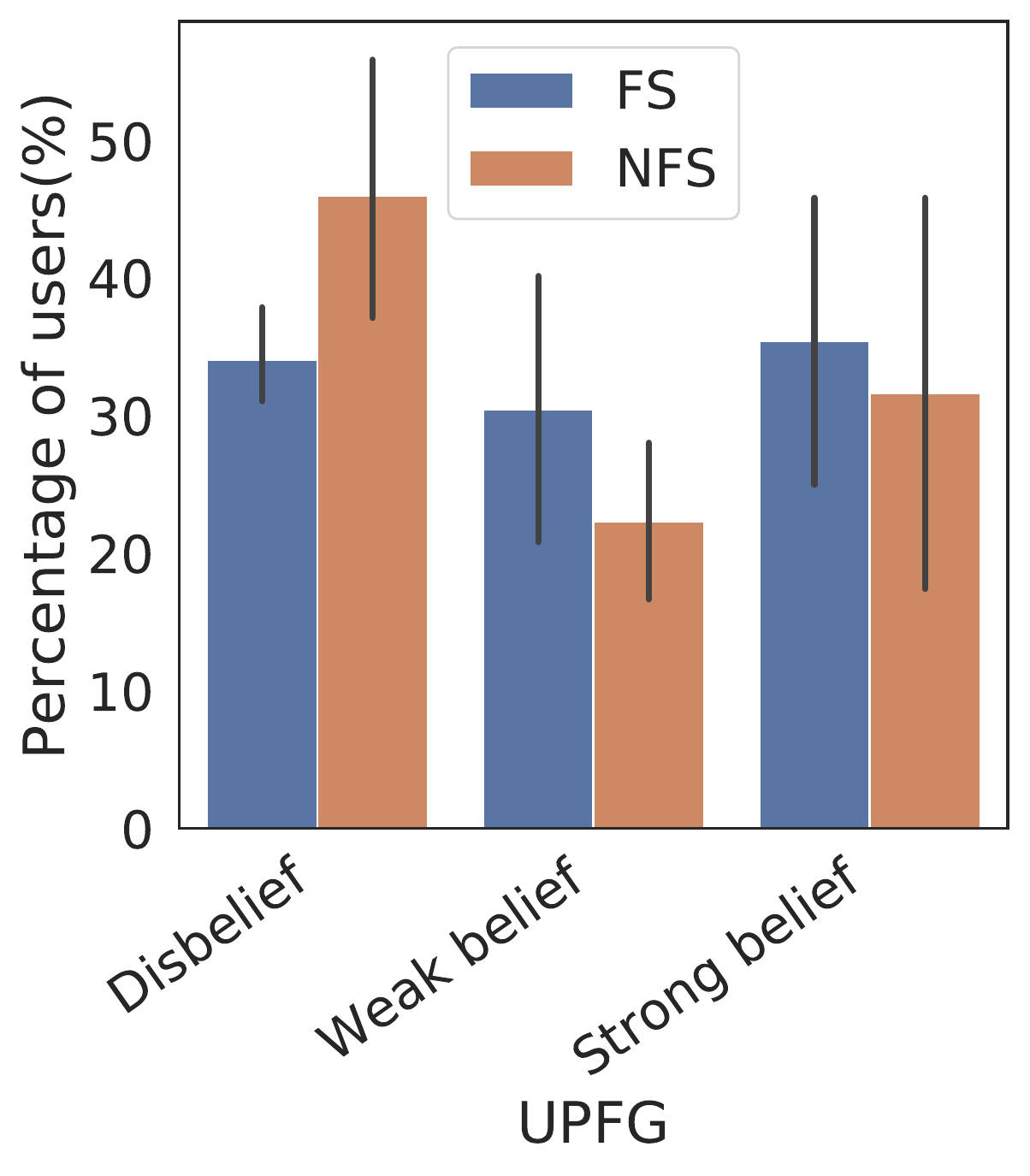}
        \label{fig:belief_fs}
    \end{minipage}%
    \begin{minipage}{.155\textwidth}
        \includegraphics[width=\linewidth]{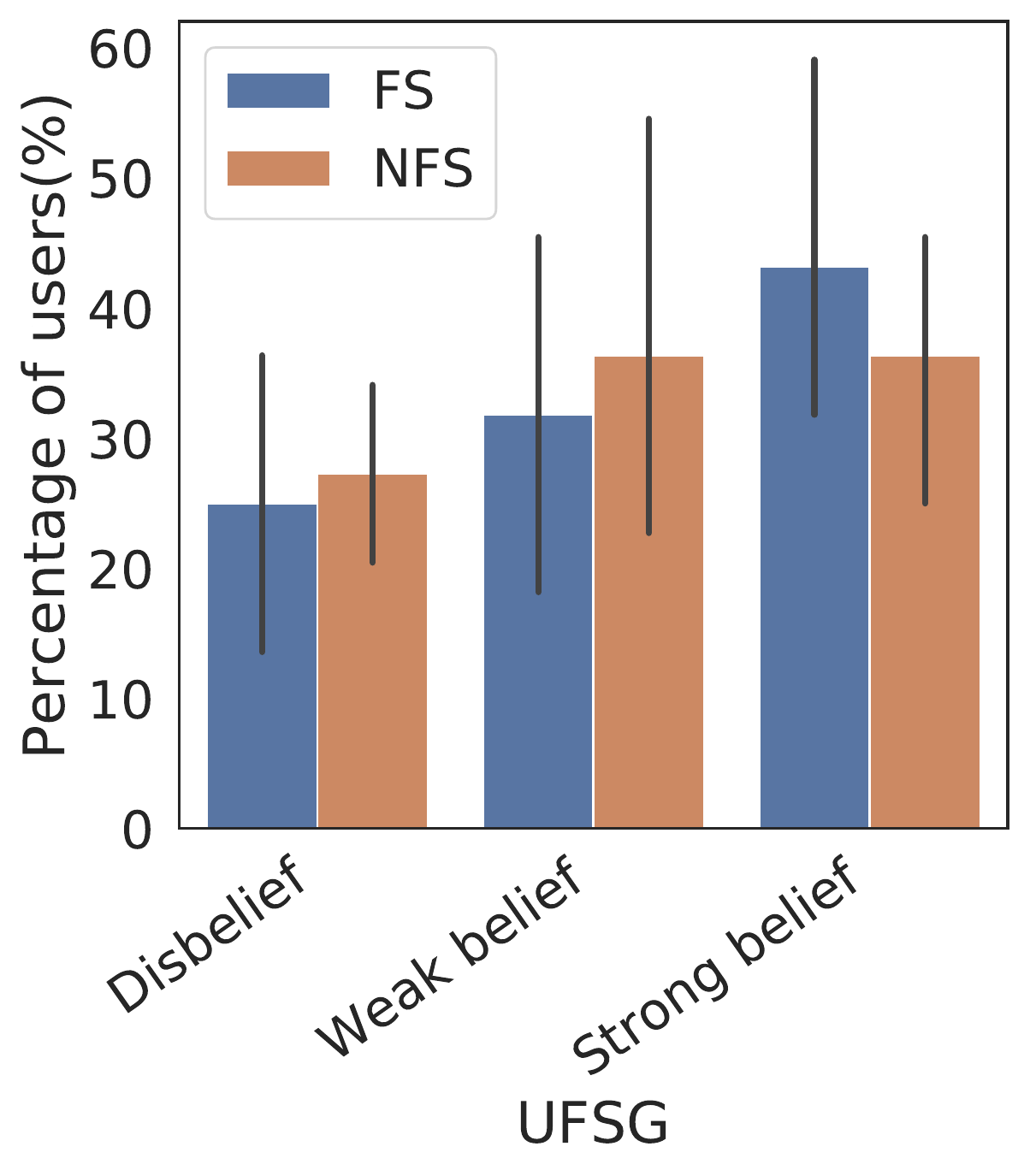}
        \label{fig:belief_ufsg}
    \end{minipage}%
    \begin{minipage}{.155\textwidth}
        \includegraphics[width=\linewidth]{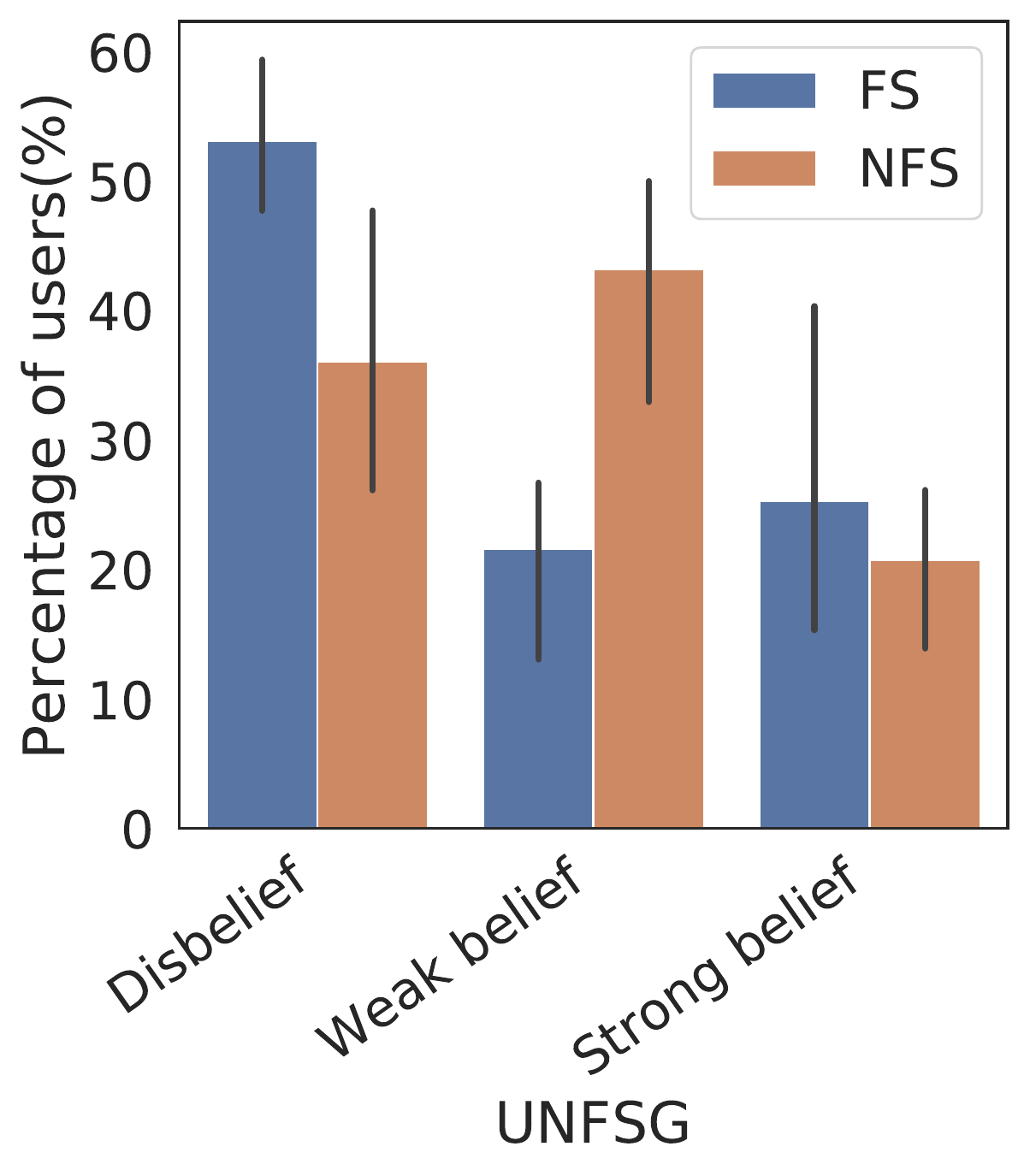}
        \label{fig:belief_unfsg}
    \end{minipage}%
    \caption{\label{fig:belief} Belief toward fear speech (FS) and non fear speech (NFS) of users in the set (i) users posting fear speech (UPFG), (ii) users in fear speech groups (UFSG), and (iii) users in non fear speech groups. Error bars show 95\% confidence intervals.}
\vspace{-\baselineskip}
\end{figure}

\begin{figure}[!htpb]
    \centering
    \begin{minipage}{.155\textwidth}
        \includegraphics[width=\linewidth]{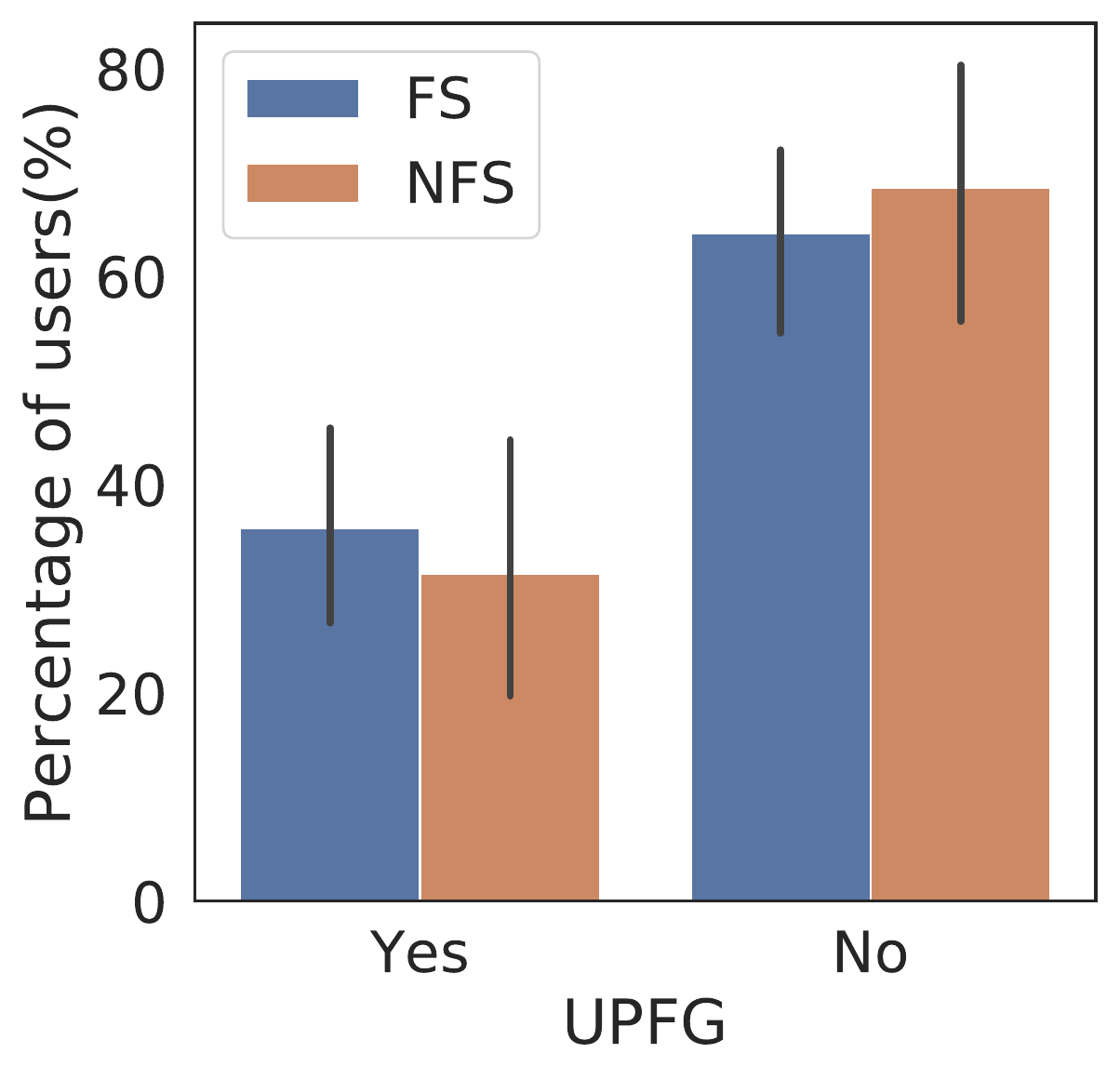}
        \label{fig:sharing_fs}
    \end{minipage}%
    \begin{minipage}{.155\textwidth}
        \centering
        \includegraphics[width=\linewidth]{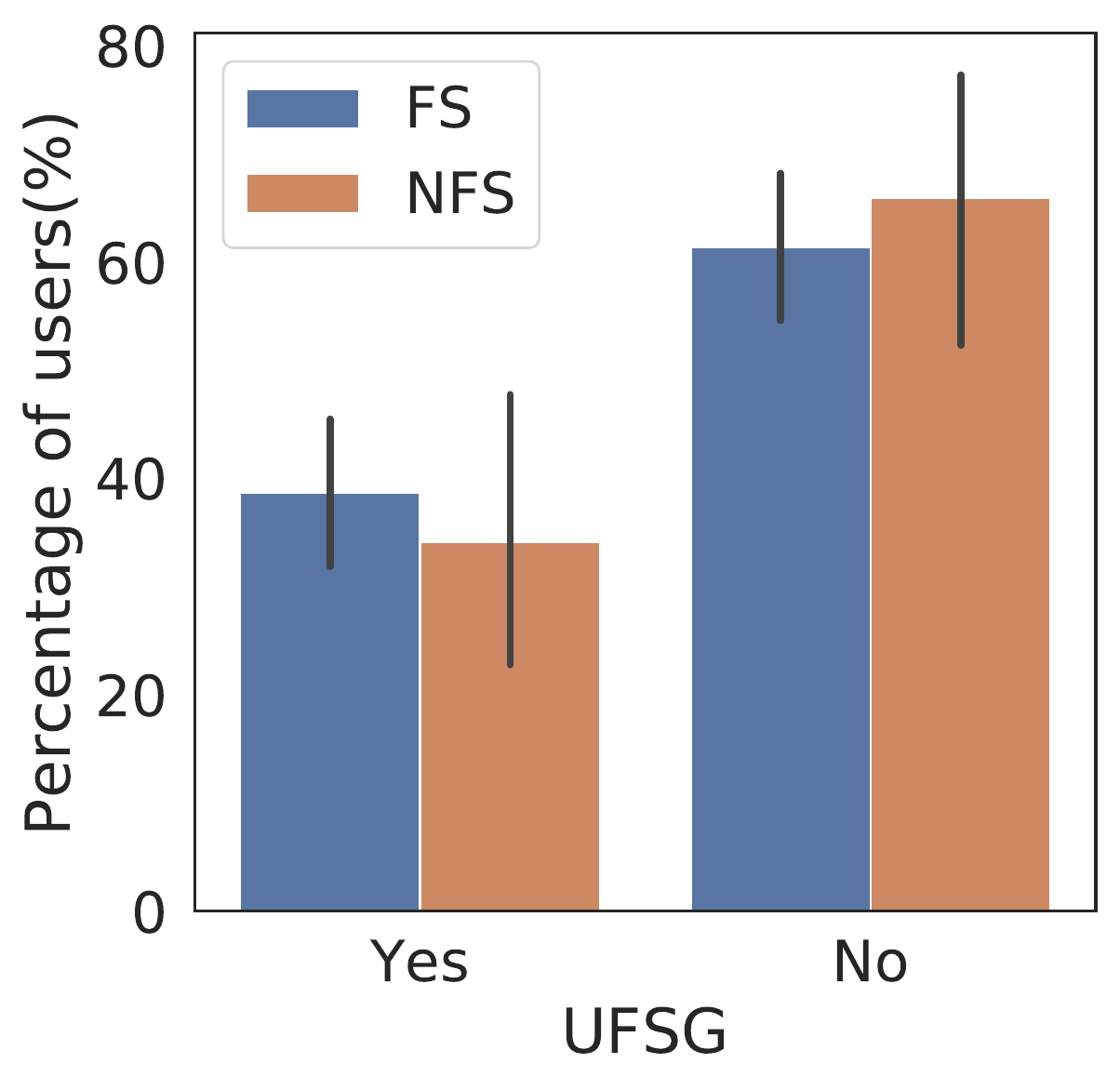}
        \label{fig:sharing_ufsg}
    \end{minipage}
    \begin{minipage}{.155\textwidth}
        \centering
        \includegraphics[width=\linewidth]{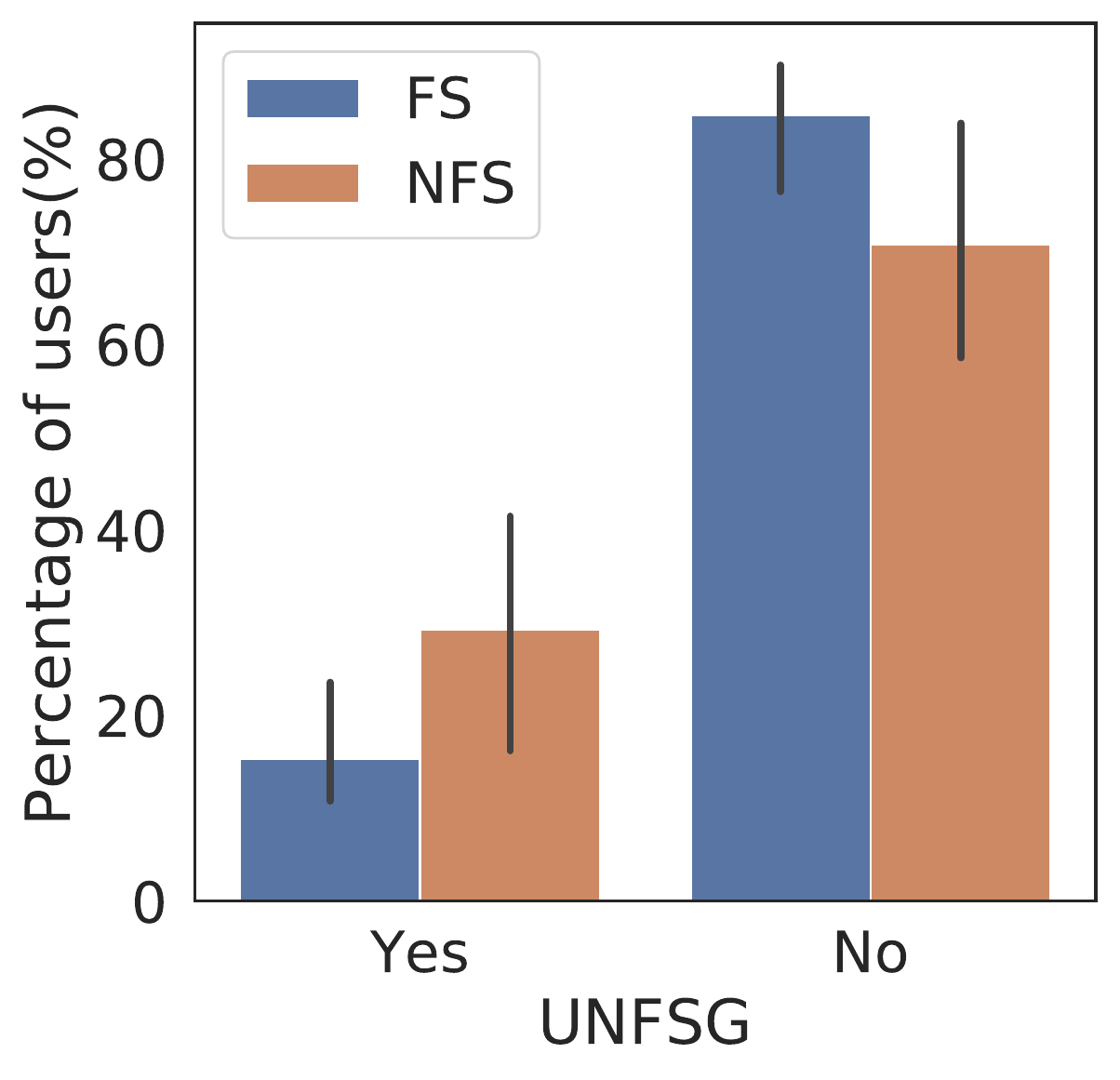}
        \label{fig:sharing_unfsg}
    \end{minipage}
    \caption{\label{fig:sharing} Sharing propensity for fear speech (FS) and non fear speech (NFS) by the users in the set (i) users posting fear speech (UPFG), (ii) users in fear speech groups (UFSG), and (iii) users in non fear speech groups.} %
\vspace{-\baselineskip}
\end{figure}

We begin with analyzing the results of the survey based on the beliefs about fear speech statements.
Figures~\ref{fig:belief} shows the beliefs of the three groups of users for the two types of statements containing fear speech (FS) and not containing fear speech (NFS).
We see that users belonging to \upfg and \ufsg have a higher probability of either weakly or strongly believing fear speech statements than non-fear speech statements. The trends are reversed when looking at the \unfsg set.
Similarly, Figure~\ref{fig:sharing} shows trends for whether the users will share statements containing fear speech or not and it clearly shows that users in \upfg and \ufsg are more likely to share fear speech.
Note that due to the low sample size, the results are not statistically significant and hence, no causal claim can be made.
However, the trends in multiple user sets show some evidence that users getting exposed are consistently more likely to believe and share fear speech statements. Further analysis on a larger sample might help attain a significant difference.

Finally, we also looked at baseline questions on the beliefs about issues related to Muslims in India conditioning on the group of the users. The results are shown in Table~\ref{tab:survey_questions1}. 
We see clear evidence that the users who belong to \upfg and \ufsg are significantly more likely to support the right wing party in power (BJP), blame Muslims for the COVID-19 hotspot in Nizamuddin Markaz, and to support the Citizenship Ammendment Bill. There is no consistent trend between users who are just a part of a group where fear speech is posted vs. users who post fear speech.

\begin{table}[]
\caption{Support for various questions given the type of users. We see that users from \upfg and \ufsg are much more likely to support anti-Muslim issues.}
\label{tab:survey_questions1}
\small
\begin{tabular}{llll}
                                                                  & \upfg & \ufsg & \unfsg \\
                                                \hline
Support BJP                                                       & 43.58\%              & 27.27\%              & 15.78\%               \\
\hline
\begin{tabular}[c]{@{}l@{}}Blame Muslims\\ for COVID\end{tabular} & 25.64\%              & 31.81\%              & 15.28\%               \\
\hline
Support CAB                                                       & 70.51\%              & 90.91\%              & 42.10\%   \\           
\hline\\
\end{tabular}
\vspace{-\baselineskip}
\end{table}

Even though the response rate to our survey was small, the overall paradigm of being able to create and launch  surveys conditioned on prior observational data is quite powerful and can be useful in providing valuable insights complementing many social media datasets.

\subsection{Summary of insights}

Overall, the analysis in this section reveals several insights into the usage of fear speech in Indian WhatsApp public groups. Our analysis proceeded along two dimensions: content and users.

We observed that the fear speech messages have a higher spread and larger lifetime when compared to non fear speech messages. The fear speech messages talk about topics such as `aggression', `crime', `hate', `fighting' and `negative emotions' in general. Using topic modeling, we found that there are concerted narratives which drive fear speech, focused on already debunked conspiracy theories showcasing Muslims to be criminals and Hindus to be victims. We showcased the prevalence and use of various emojis to emphasize the several aspects of the message and dehumanize Muslims. 
Finally, when compared to hate speech, fear speech is found to be significantly less toxic. 

We then looked users who posted fear speech messages and found that these users are popular and occupy central positions in the network, which in part explains the popularity of the fear speech content, allowing them to disseminate such messages much more easily.
Using a survey of these users, we show that fear speech users are more likely to believe and share fear speech related statements and significantly believe or support in anti-Muslim issues.

\section{Automatic Fear Speech Detection}
In this section, we develop models for automatic detection of fear speech.
We tested a wide range of models for our use case.

\noindent\textbf{Classical models}. We first used Doc2Vec embeddings~\cite{le2014distributed} with 100 dimensional vectors to represent a post. We use Logistic Regression and SVM with RBF kernel as the classifier.

\noindent\textbf{LASER-LSTM}: In these, we decomposed the paragraphs into sentences using a multilingual sentence tokenizer~\cite{johnsonetal2014}. Then we used LASER embeddings to represent the sentences~\cite{artetxe2019massively} which produces a sentence embedding of 1024 dimension per sentence. These sequences of sentence representations were then passed through a LSTM~\cite{hochreiter1997long} model to get a document level representation. Then fully connected layer was used to train the model. For the LSTM, we set hidden layer dimension to 128 and used the Adam optimizer~\cite{kingma2014adam} with a learning rate of 0.01. %

\noindent\textbf{Transformers}: Transformers are a recent NLP architecture which are formed using a stack of self-attention blocks. There are two models in the transformers, which can handle multilingual posts -- multilingualBERT\footnote{We use the \textit{multilingual-bert-base-cased} model having 12-layer, 768-hidden, 12-heads, 110M parameters trained on 104 languages from Wikipedia}~\cite{devlin2018bert} and XLM-Roberta\footnote{
we use \textit{xlm-roberta-base} model  having ~125M parameters with 12-layers, 768-hidden-state, 3072 feed-forward hidden-state, 8-heads trained on 100 languages from Common Crawl.} \cite{ruder2019unsupervised}. Both of them are limited in the number of tokens they can handle (512 at max). We set the number of tokens $n=256$ for all the experiments due to system limitations. We used tokens from different parts of the sentences for the classification task \cite{adhikari2019docbert}; these are (a) $n$-tokens from the start, (b) $n$-tokens from the end, and, (c) ($\frac{n}{2}$)-tokens from the start and ($\frac{n}{2}$)-tokens from the end append together by a <SEP> token. For optimization, Adam optimizer~\cite{kingma2014adam} was used with a learning rate of 2e-5.%

All the results are reported using the 5-fold cross validation. In each fold, we train on the 4 splits, use 1 split for validation. The details of the performance of various models on the validation set are reported in  Table~\ref{tab:model_results}.  We select the top performing model --- XLM-Roberta+LR ($5^\textrm{th}$ row) based on the AUC-ROC score (0.83). This metric is deemed effective in many of the past works~\cite{PMID:31717760,salminen2020developing}, as it is not affected by the threshold. In order to get an idea of the prevalence of fear speech in our dataset we used the best model and ran inference on the posts having Muslim keywords. In total, we got around 18k fear speech out of which \~12k were unique. Since the model was not very precise (precision 0.51) when detecting the fear speech class, we did not proceed with further analysis on predicted data. The lower precision of the most advanced NLP models (while detecting fear speech), leaves ample scope for future research.

\begin{table*}[!htpb]
\centering
\caption{Model performance on the task of classification of fear speech. For each column \textit{best performance} is shown in bold and the \textit{second best} is underlined.}
\label{tab:model_results}
\small
\begin{tabular}{llllll}
\textbf{Model}              & \textbf{Features} & \textbf{Accuracy} & \textbf{F1-Macro} & \textbf{AUC-ROC} & \textbf{Precision (FS)} \\\hline
Logistic regression         & Doc2Vec                            & 0.72     & 0.65     & 0.74    & 0.44 \\\hline
SVC (with Rbf kernel)       & Doc2vec                            & \underline{0.75}     &\underline{0.69} & 0.77 & \underline{0.49}   \\
LSTM                        & LASER embeddings                   & 0.66     & 0.63     & 0.76    & 0.39\\\hline
XLM-Roberta + LR            & Raw text (first 256 tokens)        & 0.70     & 0.65     & \underline{0.82} & 0.42   \\
XLM-Roberta + LR            & Raw text (first 128  and last 128) & \textbf{0.76}     & \textbf{0.71}    & \textbf{0.83} & \textbf{0.51}  \\
XLM-Roberta + LR            & Raw text (last 256 tokens)         & 0.72     & 0.68     & 0.81    & 0.45 \\
mBERT + LR                  & Raw text (first 256 tokens)        & 0.70     & 0.65     & 0.80    & 0.45\\  
mBERT + LR                  & Raw text (first 128  and last 128) & 0.72     & 0.65     & 0.80    & 0.48 \\
mBERT + LR                  & Raw text (last 256 tokens)         & 0.67     & 0.63     & 0.79    & 0.42\\\hline

\end{tabular}

\end{table*}

In order to investigate further, we extract all the data points where the model's prediction was wrong. Then we randomly sampled 200 examples from this set and passed them through LIME \cite{ribeiro2016should} --- a model explanation toolkit. Each post was passed through LIME. LIME returns the top words which affected the classification. We observed this top words per post to identify if there exist any prominent patterns among the wrong predictions. We noted two important patterns, which were recurrent --- (a) the model was predicting the wrong class based on confounding factors (CF) for, e.g., stop words in Hindi/English, (b) in few other cases, the models were able to base their predictions on the target (Muslims) but failed to capture the emotion of fear, i.e., target but no emotion (TNE). We noted examples showing each of these types in Table \ref{tab:example_errors}.

\begin{table}[!tbh]
\centering
\caption{Examples where the model could not predict the ground truth (GT) and type of error that happened (CF or TNE). We also show the top words (highlighted in yellow) which affected the classification (as returned by LIME).}
\footnotesize
\begin{tabular}{p{6cm}l}
\textbf{Text (translated from Hindi)} & \textbf{GT and TE} \\\hline Big breaking: Dharmendra Shinde, a marriageist Hindu brother, was murdered by Muslims for taking out the procession of Dalit \hl{in front of} the mosque \dots Mr. Manoj Parmar \hl{reached} Piplarawa and showed communal harmony and prevented the \hl{atmosphere} from deteriorating as well as taking appropriate action against the accused \dots \hl{There} is still time for all Hindus to stay organized, otherwise, India will not take much time to become Pakistan. Share it as much as the media is not showing it " & FS and CF\\\hline
Increase brotherhood, in the last journey, on uttering the name of Ram, the \hl{procession} was attacked mercilessly by the \hl{Muslims}. This was the only thing \hl{remaining} to happen to \hl{Hindus}, now that has also happened & FS and TNE\\
\end{tabular}
\label{tab:example_errors}
\vspace{-\baselineskip}
\end{table}

\section{Discussion}%

In this paper, using a large dataset of public WhatsApp conversations from India, we perform the first large scale quantitative study to characterize fear speech.
First, we manually annotated a large fear speech dataset. 
Analyzing the annotated dataset revealed certain peculiar aspects about the content of the fear speech as well as the users who post them.
We observe that the fear speech messages are re-posted by more number of users to more groups as compared to non fear speech messages, primarily because the users who post such messages are centrally placed in the WhatsApp network. 
Fear speech messages clearly fit into a set of topics relating to aggression, crime, and, violence showcasing Muslims as criminals and using dehumanizing representations of them. 
We utilized state-of-the-art NLP models to develop classification models for fear speech detection and show that the best performing model can not be reliably used to classify fear speech automatically. Using Facebook ads survey, we observe that the fear speech users are more likely to believe and share fear speech related statements and significantly believe or support in anti-Muslim issues.

Given the prevalence of WhatsApp in India (and in the global south in general), the problem of fear speech and the advances in understanding its characteristics and prevalence are valuable.
This is especially pertinent since WhatsApp is an end-to-end encrypted platform, 
where content moderation completely left out to the users. 
In our qualitative analysis of fear speech messages, we observed that many of them are based on factually inaccurate information meant to mislead the reader.
Most users are either not equipped with the know-how to detect such inaccuracies or are not interested, hence getting more and more entrenched in dangerous beliefs about a community.
We hope our paper will help begin a conversation on the importance of understanding and monitoring such dangerous speech. 

One of the solutions to countering and reducing dangerous speech given the current end-to-end encrypted model on WhatsApp is to educate the users on the facts and encouraging with-in community discussion. 
Even though it is rare, we found some cases where the fear speech was countered with some positive speech. An example message -- \textit{``The biggest challenge facing Indian Muslims at this time is how to prove themselves as patriots? From media to social media, Muslims are under siege, IT cell has waged a complete war against Muslims. \dots A very deep conspiracy is going on to break the country but we have to work hard to connect the country''}. Such counter messages could be helpful in mitigating the spread of fear speech. Identifying users who post such messages and providing them incentives might be a good way forward.

Developing a client-side classifier, which can reside on a users device might be another option. More research needs to be done on both the accuracy of the model and the ability to compress it to fit on a smartphone.
Another option is for platforms to make use of data from open social networks like Facebook to train fear speech models which can later be applied to closed platforms like WhatsApp.

Though we focused on fear speech against Muslims and on WhatsApp in this paper, we can clearly see that the scope of this problem is neither limited to Muslims nor to WhatsApp. 
Our analysis also revealed instances of fear speech against other communities as well and as previous qualitative research suggests, the problem of fear speech might have a global context~\cite{klein2017fanaticism}. 
A quick Google search of our fear speech messages revealed the prevalence of the fear speech messages on other platforms, such as Facebook\footnote{e.g. \url{https://www.facebook.com/The.Sanatana.Dharma/posts/2740274036014148}} and YouTube\footnote{e.g. \url{https://www.youtube.com/watch?v=j1U9JwNsklU}}.
We hope that the dataset we release from our work will allow the community to build upon our findings and extend the quantitative research on fear speech broadly.

The focus of this paper was to introduce fear speech as an important and distinct topic to the Computational Social science audience at the Web Conference, and encourage a quantitative analysis of fear speech. 
However, there are a lot of fundamental similarities in fear speech with prior work on hate speech. 
Efforts should be made to understand these similarities and build datasets and analyses that encompass such broader versions of dangerous speech.

\noindent\textbf{Limitations}. As with any empirical work, this study has its limitations. 
First, the data used is a convenience sample of public WhatsApp discussions on politics. 
Given that WhatsApp does not provide an API or tools to access the data, there is no way of knowing the representativeness of our dataset. 
This should be kept in mind while interpreting the results. However, we have been careful through out the paper stressing that this is a convenience sample, and that our objective was to focus on the problem of fear speech rather than the representativeness of our results on all of WhatsApp.

\bibliographystyle{ACM-Reference-Format}
\bibliography{references.bib}


\begin{thebibliography}{69}


\ifx \showCODEN    \undefined \def \showCODEN     #1{\unskip}     \fi
\ifx \showDOI      \undefined \def \showDOI       #1{#1}\fi
\ifx \showISBNx    \undefined \def \showISBNx     #1{\unskip}     \fi
\ifx \showISBNxiii \undefined \def \showISBNxiii  #1{\unskip}     \fi
\ifx \showISSN     \undefined \def \showISSN      #1{\unskip}     \fi
\ifx \showLCCN     \undefined \def \showLCCN      #1{\unskip}     \fi
\ifx \shownote     \undefined \def \shownote      #1{#1}          \fi
\ifx \showarticletitle \undefined \def \showarticletitle #1{#1}   \fi
\ifx \showURL      \undefined \def \showURL       {\relax}        \fi
\providecommand\bibfield[2]{#2}
\providecommand\bibinfo[2]{#2}
\providecommand\natexlab[1]{#1}
\providecommand\showeprint[2][]{arXiv:#2}

\bibitem[\protect\citeauthoryear{??}{per}{2019}]%
        {perspective}
 \bibinfo{year}{2019}\natexlab{}.
\newblock \bibinfo{title}{Perspective API}.
\newblock \bibinfo{howpublished}{\url{https://www.perspectiveapi.com/}}.
\newblock


\bibitem[\protect\citeauthoryear{Adhikari, Ram, Tang, and Lin}{Adhikari
  et~al\mbox{.}}{2019}]%
        {adhikari2019docbert}
\bibfield{author}{\bibinfo{person}{Ashutosh Adhikari}, \bibinfo{person}{Achyudh
  Ram}, \bibinfo{person}{Raphael Tang}, {and} \bibinfo{person}{Jimmy Lin}.}
  \bibinfo{year}{2019}\natexlab{}.
\newblock \showarticletitle{Docbert: Bert for document classification}.
\newblock \bibinfo{journal}{\emph{arXiv preprint arXiv:1904.08398}}
  (\bibinfo{year}{2019}).
\newblock


\bibitem[\protect\citeauthoryear{Aluru, Mathew, Saha, and Mukherjee}{Aluru
  et~al\mbox{.}}{2020}]%
        {aluru2020deep}
\bibfield{author}{\bibinfo{person}{Sai~Saket Aluru}, \bibinfo{person}{Binny
  Mathew}, \bibinfo{person}{Punyajoy Saha}, {and} \bibinfo{person}{Animesh
  Mukherjee}.} \bibinfo{year}{2020}\natexlab{}.
\newblock \showarticletitle{Deep Learning Models for Multilingual Hate Speech
  Detection}.
\newblock \bibinfo{journal}{\emph{arXiv preprint arXiv:2004.06465}}
  (\bibinfo{year}{2020}).
\newblock


\bibitem[\protect\citeauthoryear{Artetxe and Schwenk}{Artetxe and
  Schwenk}{2019}]%
        {artetxe2019massively}
\bibfield{author}{\bibinfo{person}{Mikel Artetxe} {and} \bibinfo{person}{Holger
  Schwenk}.} \bibinfo{year}{2019}\natexlab{}.
\newblock \showarticletitle{Massively multilingual sentence embeddings for
  zero-shot cross-lingual transfer and beyond}.
\newblock \bibinfo{journal}{\emph{Transactions of the Association for
  Computational Linguistics}}  \bibinfo{volume}{7} (\bibinfo{year}{2019}),
  \bibinfo{pages}{597--610}.
\newblock


\bibitem[\protect\citeauthoryear{Arun}{Arun}{2019}]%
        {arun2019whatsapp}
\bibfield{author}{\bibinfo{person}{Chinmayi Arun}.}
  \bibinfo{year}{2019}\natexlab{}.
\newblock \showarticletitle{On WhatsApp, Rumours, and Lynchings}.
\newblock \bibinfo{journal}{\emph{Economic \& Political Weekly}}
  \bibinfo{volume}{54}, \bibinfo{number}{6} (\bibinfo{year}{2019}),
  \bibinfo{pages}{30--35}.
\newblock


\bibitem[\protect\citeauthoryear{Attwell}{Attwell}{2018}]%
        {Religiou26:online}
\bibfield{author}{\bibinfo{person}{Robert Attwell}.}
  \bibinfo{year}{2018}\natexlab{}.
\newblock \bibinfo{title}{Religious Intolerance, Lynch Mobs and Social Media in
  India | GSI}.
\newblock
  \bibinfo{howpublished}{\url{https://gsi.s-rminform.com/articles/religious-intolerance-lynch-mobs-and-social-media-in-india}}.
\newblock
\newblock
\shownote{(Accessed on 01/22/2021).}


\bibitem[\protect\citeauthoryear{Basile, Bosco, Fersini, Nozza, Patti,
  Rangel~Pardo, Rosso, and Sanguinetti}{Basile et~al\mbox{.}}{2019}]%
        {basile-etal-2019-semeval}
\bibfield{author}{\bibinfo{person}{Valerio Basile}, \bibinfo{person}{Cristina
  Bosco}, \bibinfo{person}{Elisabetta Fersini}, \bibinfo{person}{Debora Nozza},
  \bibinfo{person}{Viviana Patti}, \bibinfo{person}{Francisco~Manuel
  Rangel~Pardo}, \bibinfo{person}{Paolo Rosso}, {and} \bibinfo{person}{Manuela
  Sanguinetti}.} \bibinfo{year}{2019}\natexlab{}.
\newblock \showarticletitle{{S}em{E}val-2019 Task 5: Multilingual Detection of
  Hate Speech Against Immigrants and Women in {T}witter}. In
  \bibinfo{booktitle}{\emph{Proceedings of the 13th International Workshop on
  Semantic Evaluation}}. \bibinfo{publisher}{Association for Computational
  Linguistics}, \bibinfo{pages}{54--63}.
\newblock


\bibitem[\protect\citeauthoryear{Basu}{Basu}{2019}]%
        {Manufact87:online}
\bibfield{author}{\bibinfo{person}{Soma Basu}.}
  \bibinfo{year}{2019}\natexlab{}.
\newblock \bibinfo{title}{Manufacturing Islamophobia on WhatsApp in India –
  The Diplomat}.
\newblock
  \bibinfo{howpublished}{\url{https://thediplomat.com/2019/05/manufacturing-islamophobia-on-whatsapp-in-india}}.
\newblock
\newblock
\shownote{(Accessed on 01/22/2021).}


\bibitem[\protect\citeauthoryear{Benesch}{Benesch}{2012}]%
        {benesch2012dangerous}
\bibfield{author}{\bibinfo{person}{Susan Benesch}.}
  \bibinfo{year}{2012}\natexlab{}.
\newblock \showarticletitle{Dangerous speech: A proposal to prevent group
  violence}.
\newblock  (\bibinfo{year}{2012}).
\newblock


\bibitem[\protect\citeauthoryear{Bisht and Naqvi}{Bisht and Naqvi}{2020}]%
        {HowTabli17:online}
\bibfield{author}{\bibinfo{person}{Akash Bisht} {and} \bibinfo{person}{Sadiq
  Naqvi}.} \bibinfo{year}{2020}\natexlab{}.
\newblock \bibinfo{title}{How Tablighi Jamaat event became India’s worst
  coronavirus vector | Coronavirus pandemic News | Al Jazeera}.
\newblock
  \bibinfo{howpublished}{\url{https://www.aljazeera.com/news/2020/4/7/how-tablighi-jamaat-event-became-indias-worst-coronavirus-vector}}.
\newblock
\newblock
\shownote{(Accessed on 01/22/2021).}


\bibitem[\protect\citeauthoryear{Bisri}{Bisri}{2020}]%
        {bisri2020indonesian}
\bibfield{author}{\bibinfo{person}{Hasan Bisri}.}
  \bibinfo{year}{2020}\natexlab{}.
\newblock \showarticletitle{The Indonesian-Moderate Muslim Communities Opinion
  on Social Media Hate Speech}.
\newblock \bibinfo{journal}{\emph{International Journal of Psychosocial
  Rehabilitation}} \bibinfo{volume}{24}, \bibinfo{number}{8}
  (\bibinfo{year}{2020}), \bibinfo{pages}{10941--10951}.
\newblock


\bibitem[\protect\citeauthoryear{Bj{\o}rnskov and Mchangama}{Bj{\o}rnskov and
  Mchangama}{2019}]%
        {bjornskov2019social}
\bibfield{author}{\bibinfo{person}{Christian Bj{\o}rnskov} {and}
  \bibinfo{person}{Jacob Mchangama}.} \bibinfo{year}{2019}\natexlab{}.
\newblock \showarticletitle{Do Social Rights Affect Social Outcomes?}
\newblock \bibinfo{journal}{\emph{American Journal of Political Science}}
  \bibinfo{volume}{63}, \bibinfo{number}{2} (\bibinfo{year}{2019}),
  \bibinfo{pages}{452--466}.
\newblock


\bibitem[\protect\citeauthoryear{Blondel, Guillaume, Lambiotte, and
  Lefebvre}{Blondel et~al\mbox{.}}{2008}]%
        {blondel2008fast}
\bibfield{author}{\bibinfo{person}{Vincent~D Blondel},
  \bibinfo{person}{Jean-Loup Guillaume}, \bibinfo{person}{Renaud Lambiotte},
  {and} \bibinfo{person}{Etienne Lefebvre}.} \bibinfo{year}{2008}\natexlab{}.
\newblock \showarticletitle{Fast unfolding of communities in large networks}.
\newblock \bibinfo{journal}{\emph{Journal of statistical mechanics: theory and
  experiment}} \bibinfo{volume}{2008}, \bibinfo{number}{10}
  (\bibinfo{year}{2008}), \bibinfo{pages}{P10008}.
\newblock


\bibitem[\protect\citeauthoryear{Bohra, Vijay, Singh, Akhtar, and
  Shrivastava}{Bohra et~al\mbox{.}}{2018}]%
        {bohra-etal-2018-dataset}
\bibfield{author}{\bibinfo{person}{Aditya Bohra}, \bibinfo{person}{Deepanshu
  Vijay}, \bibinfo{person}{Vinay Singh}, \bibinfo{person}{Syed~Sarfaraz
  Akhtar}, {and} \bibinfo{person}{Manish Shrivastava}.}
  \bibinfo{year}{2018}\natexlab{}.
\newblock \showarticletitle{A Dataset of {H}indi-{E}nglish Code-Mixed Social
  Media Text for Hate Speech Detection}. In
  \bibinfo{booktitle}{\emph{Proceedings of the Second Workshop on Computational
  Modeling of People{'}s Opinions, Personality, and Emotions in Social Media}}.
  \bibinfo{publisher}{Association for Computational Linguistics},
  \bibinfo{pages}{36--41}.
\newblock


\bibitem[\protect\citeauthoryear{Buyse}{Buyse}{2014}]%
        {buyse2014words}
\bibfield{author}{\bibinfo{person}{Antoine Buyse}.}
  \bibinfo{year}{2014}\natexlab{}.
\newblock \showarticletitle{Words of violence: Fear speech, or how violent
  conflict escalation relates to the freedom of expression}.
\newblock \bibinfo{journal}{\emph{Hum. Rts. Q.}}  \bibinfo{volume}{36}
  (\bibinfo{year}{2014}), \bibinfo{pages}{779}.
\newblock


\bibitem[\protect\citeauthoryear{Caetano, de~Oliveira, Lima, Marques-Neto,
  Magno, Meira~Jr, and Almeida}{Caetano et~al\mbox{.}}{2018}]%
        {caetano2018analyzing}
\bibfield{author}{\bibinfo{person}{Josemar~Alves Caetano},
  \bibinfo{person}{Jaqueline~Faria de Oliveira},
  \bibinfo{person}{H{\'e}lder~Seixas Lima}, \bibinfo{person}{Humberto~T
  Marques-Neto}, \bibinfo{person}{Gabriel Magno}, \bibinfo{person}{Wagner
  Meira~Jr}, {and} \bibinfo{person}{Virg{\'\i}lio~AF Almeida}.}
  \bibinfo{year}{2018}\natexlab{}.
\newblock \showarticletitle{Analyzing and characterizing political discussions
  in WhatsApp public groups}.
\newblock \bibinfo{journal}{\emph{arXiv preprint arXiv:1804.00397}}
  (\bibinfo{year}{2018}).
\newblock


\bibitem[\protect\citeauthoryear{Chung, Kuzmenko, Tekiroglu, and Guerini}{Chung
  et~al\mbox{.}}{2019}]%
        {chung2019conan}
\bibfield{author}{\bibinfo{person}{Yi-Ling Chung}, \bibinfo{person}{Elizaveta
  Kuzmenko}, \bibinfo{person}{Serra~Sinem Tekiroglu}, {and}
  \bibinfo{person}{Marco Guerini}.} \bibinfo{year}{2019}\natexlab{}.
\newblock \showarticletitle{CONAN-COunter NArratives through Nichesourcing: a
  Multilingual Dataset of Responses to Fight Online Hate Speech}. In
  \bibinfo{booktitle}{\emph{57th Annual Meeting of the Association for
  Computational Linguistics}}. \bibinfo{pages}{2819--2829}.
\newblock


\bibitem[\protect\citeauthoryear{Del~Vigna, Cimino, Dell’Orletta, Petrocchi,
  and Tesconi}{Del~Vigna et~al\mbox{.}}{2017}]%
        {del2017hate}
\bibfield{author}{\bibinfo{person}{Fabio Del~Vigna}, \bibinfo{person}{Andrea
  Cimino}, \bibinfo{person}{Felice Dell’Orletta}, \bibinfo{person}{Marinella
  Petrocchi}, {and} \bibinfo{person}{Maurizio Tesconi}.}
  \bibinfo{year}{2017}\natexlab{}.
\newblock \showarticletitle{Hate me, hate me not: Hate speech detection on
  facebook}. In \bibinfo{booktitle}{\emph{In Proceedings of the First Italian
  Conference on Cybersecurity (ITASEC17)}}.
\newblock


\bibitem[\protect\citeauthoryear{Devlin, Chang, Lee, and Toutanova}{Devlin
  et~al\mbox{.}}{2019}]%
        {devlin2018bert}
\bibfield{author}{\bibinfo{person}{Jacob Devlin}, \bibinfo{person}{Ming-Wei
  Chang}, \bibinfo{person}{Kenton Lee}, {and} \bibinfo{person}{Kristina
  Toutanova}.} \bibinfo{year}{2019}\natexlab{}.
\newblock \showarticletitle{BERT: Pre-training of Deep Bidirectional
  Transformers for Language Understanding}. In
  \bibinfo{booktitle}{\emph{Proceedings of the 2019 Conference of the NAACL}}.
  \bibinfo{pages}{4171--4186}.
\newblock


\bibitem[\protect\citeauthoryear{ElSherief, Kulkarni, Nguyen, Wang, and
  Belding}{ElSherief et~al\mbox{.}}{2018}]%
        {elsherief2018hate}
\bibfield{author}{\bibinfo{person}{Mai ElSherief}, \bibinfo{person}{Vivek
  Kulkarni}, \bibinfo{person}{Dana Nguyen}, \bibinfo{person}{William~Yang
  Wang}, {and} \bibinfo{person}{Elizabeth~M. Belding}.}
  \bibinfo{year}{2018}\natexlab{}.
\newblock \showarticletitle{Hate Lingo: {A} Target-Based Linguistic Analysis of
  Hate Speech in Social Media}.
\newblock \bibinfo{journal}{\emph{Proceedings of the Twelfth International
  Conference on Web and Social Media, {ICWSM} 2018}} (\bibinfo{year}{2018}).
\newblock


\bibitem[\protect\citeauthoryear{Fast, Chen, and Bernstein}{Fast
  et~al\mbox{.}}{2016}]%
        {fast2016empath}
\bibfield{author}{\bibinfo{person}{Ethan Fast}, \bibinfo{person}{Binbin Chen},
  {and} \bibinfo{person}{Michael~S Bernstein}.}
  \bibinfo{year}{2016}\natexlab{}.
\newblock \showarticletitle{Empath: Understanding topic signals in large-scale
  text}. In \bibinfo{booktitle}{\emph{Proceedings of the 2016 CHI Conference on
  Human Factors in Computing Systems}}. \bibinfo{pages}{4647--4657}.
\newblock


\bibitem[\protect\citeauthoryear{Fortuna and Nunes}{Fortuna and Nunes}{2018}]%
        {fortuna2018survey}
\bibfield{author}{\bibinfo{person}{Paula Fortuna} {and}
  \bibinfo{person}{S{\'e}rgio Nunes}.} \bibinfo{year}{2018}\natexlab{}.
\newblock \showarticletitle{A survey on automatic detection of hate speech in
  text}.
\newblock \bibinfo{journal}{\emph{ACM Computing Surveys (CSUR)}}
  \bibinfo{volume}{51}, \bibinfo{number}{4} (\bibinfo{year}{2018}),
  \bibinfo{pages}{1--30}.
\newblock


\bibitem[\protect\citeauthoryear{Gagliardone}{Gagliardone}{2019}]%
        {gagliardone2019extreme}
\bibfield{author}{\bibinfo{person}{Iginio Gagliardone}.}
  \bibinfo{year}{2019}\natexlab{}.
\newblock \showarticletitle{Extreme Speech| Defining Online Hate and Its
  “Public Lives”: What is the Place for “Extreme Speech”?}
\newblock \bibinfo{journal}{\emph{International Journal of Communication}}
  \bibinfo{volume}{13} (\bibinfo{year}{2019}), \bibinfo{pages}{20}.
\newblock


\bibitem[\protect\citeauthoryear{Garimella and Eckles}{Garimella and
  Eckles}{2020a}]%
        {Garimella_2020}
\bibfield{author}{\bibinfo{person}{Kiran Garimella} {and} \bibinfo{person}{Dean
  Eckles}.} \bibinfo{year}{2020}\natexlab{a}.
\newblock \showarticletitle{Images and Misinformation in Political Groups:
  Evidence from WhatsApp in India}.
\newblock \bibinfo{journal}{\emph{Harvard Kennedy School Misinformation
  Review}} (\bibinfo{date}{Jul} \bibinfo{year}{2020}).
\newblock
\urldef\tempurl%
\url{https://doi.org/10.37016/mr-2020-030}
\showDOI{\tempurl}


\bibitem[\protect\citeauthoryear{Garimella and Eckles}{Garimella and
  Eckles}{2020b}]%
        {garimella2020images}
\bibfield{author}{\bibinfo{person}{Kiran Garimella} {and} \bibinfo{person}{Dean
  Eckles}.} \bibinfo{year}{2020}\natexlab{b}.
\newblock \showarticletitle{Images and Misinformation in Political Groups:
  Evidence from WhatsApp in India}.
\newblock \bibinfo{journal}{\emph{Harvard Kennedy School Misinformation
  Review}} (\bibinfo{year}{2020}).
\newblock


\bibitem[\protect\citeauthoryear{Garimella and Tyson}{Garimella and
  Tyson}{2018}]%
        {garimella2018whatapp}
\bibfield{author}{\bibinfo{person}{Kiran Garimella} {and}
  \bibinfo{person}{Gareth Tyson}.} \bibinfo{year}{2018}\natexlab{}.
\newblock \showarticletitle{Whatapp Doc? A First Look at Whatsapp Public Group
  Data}. In \bibinfo{booktitle}{\emph{12th ICWSM}}.
\newblock


\bibitem[\protect\citeauthoryear{Gionis, Indyk, Motwani, et~al\mbox{.}}{Gionis
  et~al\mbox{.}}{1999}]%
        {gionis1999similarity}
\bibfield{author}{\bibinfo{person}{Aristides Gionis}, \bibinfo{person}{Piotr
  Indyk}, \bibinfo{person}{Rajeev Motwani}, {et~al\mbox{.}}}
  \bibinfo{year}{1999}\natexlab{}.
\newblock \showarticletitle{Similarity search in high dimensions via hashing}.
  In \bibinfo{booktitle}{\emph{Vldb}}, Vol.~\bibinfo{volume}{99}.
  \bibinfo{pages}{518--529}.
\newblock


\bibitem[\protect\citeauthoryear{Gottschalk, Greenberg, and
  Greenberg}{Gottschalk et~al\mbox{.}}{2008}]%
        {gottschalk2008islamophobia}
\bibfield{author}{\bibinfo{person}{Peter Gottschalk}, \bibinfo{person}{Gabriel
  Greenberg}, {and} \bibinfo{person}{Gary Greenberg}.}
  \bibinfo{year}{2008}\natexlab{}.
\newblock \bibinfo{booktitle}{\emph{Islamophobia: making Muslims the enemy}}.
\newblock \bibinfo{publisher}{Rowman \& Littlefield}.
\newblock


\bibitem[\protect\citeauthoryear{Haslwanter}{Haslwanter}{2016}]%
        {haslwanter2016analysis}
\bibfield{author}{\bibinfo{person}{Thomas Haslwanter}.}
  \bibinfo{year}{2016}\natexlab{}.
\newblock \showarticletitle{Analysis of Survival Times}.
\newblock In \bibinfo{booktitle}{\emph{An Introduction to Statistics with
  Python}}. \bibinfo{publisher}{Springer}, \bibinfo{pages}{175--180}.
\newblock


\bibitem[\protect\citeauthoryear{Hochreiter and Schmidhuber}{Hochreiter and
  Schmidhuber}{1997}]%
        {hochreiter1997long}
\bibfield{author}{\bibinfo{person}{Sepp Hochreiter} {and}
  \bibinfo{person}{J{\"u}rgen Schmidhuber}.} \bibinfo{year}{1997}\natexlab{}.
\newblock \showarticletitle{Long short-term memory}.
\newblock \bibinfo{journal}{\emph{Neural computation}} \bibinfo{volume}{9},
  \bibinfo{number}{8} (\bibinfo{year}{1997}), \bibinfo{pages}{1735--1780}.
\newblock


\bibitem[\protect\citeauthoryear{Hoffman, Bach, and Blei}{Hoffman
  et~al\mbox{.}}{2010}]%
        {hoffman2010online}
\bibfield{author}{\bibinfo{person}{Matthew Hoffman}, \bibinfo{person}{Francis~R
  Bach}, {and} \bibinfo{person}{David~M Blei}.}
  \bibinfo{year}{2010}\natexlab{}.
\newblock \showarticletitle{Online learning for latent dirichlet allocation}.
  In \bibinfo{booktitle}{\emph{advances in neural information processing
  systems}}.
\newblock


\bibitem[\protect\citeauthoryear{Hoffman-Pham et~al\mbox{.}}{Hoffman-Pham
  et~al\mbox{.}}{2020}]%
        {hoffman2020facebook}
\bibfield{author}{\bibinfo{person}{Katherine Hoffman-Pham} {et~al\mbox{.}}}
  \bibinfo{year}{2020}\natexlab{}.
\newblock \showarticletitle{Social Media Markets for Survey Research in
  Comparative Contexts: Facebook Users in Kenya}.
\newblock \bibinfo{journal}{\emph{Arxiv 1910.03448}} (\bibinfo{year}{2020}).
\newblock


\bibitem[\protect\citeauthoryear{Johnson, Burns, Stewart, and Cook}{Johnson
  et~al\mbox{.}}{2020}]%
        {johnsonetal2014}
\bibfield{author}{\bibinfo{person}{Kyle~P. Johnson}, \bibinfo{person}{Patrick
  Burns}, \bibinfo{person}{John Stewart}, {and} \bibinfo{person}{Todd Cook}.}
  \bibinfo{year}{2014--2020}\natexlab{}.
\newblock \bibinfo{title}{CLTK: The Classical Language Toolkit}.
\newblock
\newblock
\urldef\tempurl%
\url{https://github.com/cltk/cltk}
\showURL{%
\tempurl}


\bibitem[\protect\citeauthoryear{Kingma and Ba}{Kingma and Ba}{2015}]%
        {kingma2014adam}
\bibfield{author}{\bibinfo{person}{Diederik~P. Kingma} {and}
  \bibinfo{person}{Jimmy Ba}.} \bibinfo{year}{2015}\natexlab{}.
\newblock \showarticletitle{Adam: {A} Method for Stochastic Optimization}. In
  \bibinfo{booktitle}{\emph{3rd International Conference on Learning
  Representations, {ICLR} 2015, San Diego, CA, USA, May 7-9, 2015, Conference
  Track Proceedings}}.
\newblock


\bibitem[\protect\citeauthoryear{Kitsak, Gallos, Havlin, Liljeros, Muchnik,
  Stanley, and Makse}{Kitsak et~al\mbox{.}}{2010}]%
        {kitsak2010identification}
\bibfield{author}{\bibinfo{person}{Maksim Kitsak}, \bibinfo{person}{Lazaros~K
  Gallos}, \bibinfo{person}{Shlomo Havlin}, \bibinfo{person}{Fredrik Liljeros},
  \bibinfo{person}{Lev Muchnik}, \bibinfo{person}{H~Eugene Stanley}, {and}
  \bibinfo{person}{Hern{\'a}n~A Makse}.} \bibinfo{year}{2010}\natexlab{}.
\newblock \showarticletitle{Identification of influential spreaders in complex
  networks}.
\newblock \bibinfo{journal}{\emph{Nature physics}} \bibinfo{volume}{6},
  \bibinfo{number}{11} (\bibinfo{year}{2010}), \bibinfo{pages}{888--893}.
\newblock


\bibitem[\protect\citeauthoryear{Klein}{Klein}{2017}]%
        {klein2017fanaticism}
\bibfield{author}{\bibinfo{person}{Adam Klein}.}
  \bibinfo{year}{2017}\natexlab{}.
\newblock \bibinfo{booktitle}{\emph{Fanaticism, racism, and rage online:
  Corrupting the digital sphere}}.
\newblock \bibinfo{publisher}{Springer}.
\newblock


\bibitem[\protect\citeauthoryear{Kumar}{Kumar}{2018}]%
        {HowtheBJ16:online}
\bibfield{author}{\bibinfo{person}{Kuldeep Kumar}.}
  \bibinfo{year}{2018}\natexlab{}.
\newblock \bibinfo{title}{How the BJP – Master of Mixing Religion and
  Politics – Is Taking India for a Ride}.
\newblock
  \bibinfo{howpublished}{\url{https://thewire.in/communalism/bjp-masters-of-mixing-religion-politics}}.
\newblock
\newblock
\shownote{(Accessed on 01/22/2021).}


\bibitem[\protect\citeauthoryear{Le and Mikolov}{Le and Mikolov}{2014}]%
        {le2014distributed}
\bibfield{author}{\bibinfo{person}{Quoc Le} {and} \bibinfo{person}{Tomas
  Mikolov}.} \bibinfo{year}{2014}\natexlab{}.
\newblock \showarticletitle{Distributed representations of sentences and
  documents}. In \bibinfo{booktitle}{\emph{International conference on machine
  learning}}. \bibinfo{pages}{1188--1196}.
\newblock


\bibitem[\protect\citeauthoryear{Lenhard and Lenhard}{Lenhard and
  Lenhard}{2017}]%
        {effect_sizes}
\bibfield{author}{\bibinfo{person}{Wolfgang Lenhard} {and}
  \bibinfo{person}{Alexandra Lenhard}.} \bibinfo{year}{2017}\natexlab{}.
\newblock \showarticletitle{Computation of Effect Sizes}.
\newblock  (\bibinfo{date}{10} \bibinfo{year}{2017}).
\newblock
\urldef\tempurl%
\url{https://doi.org/10.13140/RG.2.2.17823.92329}
\showDOI{\tempurl}


\bibitem[\protect\citeauthoryear{Levin}{Levin}{2017}]%
        {levin2017moderators}
\bibfield{author}{\bibinfo{person}{S Levin}.} \bibinfo{year}{2017}\natexlab{}.
\newblock \showarticletitle{Moderators who had to view child abuse content sue
  microsoft, claiming ptsd}.
\newblock \bibinfo{journal}{\emph{The Guardian}} (\bibinfo{year}{2017}).
\newblock
\urldef\tempurl%
\url{https://www.theguardian.com/technology/2017/jan/11/microsoft-employees-child-abuse-lawsuit-ptsd}
\showURL{%
\tempurl}


\bibitem[\protect\citeauthoryear{Li, Bai, Yang, Liu, and Chen}{Li
  et~al\mbox{.}}{2018}]%
        {li2018co}
\bibfield{author}{\bibinfo{person}{Taoying Li}, \bibinfo{person}{Jie Bai},
  \bibinfo{person}{Xue Yang}, \bibinfo{person}{Qianyu Liu}, {and}
  \bibinfo{person}{Yan Chen}.} \bibinfo{year}{2018}\natexlab{}.
\newblock \showarticletitle{Co-Occurrence Network of High-Frequency Words in
  the Bioinformatics Literature: Structural Characteristics and Evolution}.
\newblock \bibinfo{journal}{\emph{Applied Sciences}} \bibinfo{volume}{8},
  \bibinfo{number}{10} (\bibinfo{year}{2018}), \bibinfo{pages}{1994}.
\newblock


\bibitem[\protect\citeauthoryear{Lokniti}{Lokniti}{2018}]%
        {lokniti2018}
\bibfield{author}{\bibinfo{person}{CSDS Lokniti}.}
  \bibinfo{year}{2018}\natexlab{}.
\newblock \bibinfo{title}{How widespread is WhatsApp's usage in India?}
\newblock
\newblock
\urldef\tempurl%
\url{https://www.livemint.com/Technology/O6DLmIibCCV5luEG9XuJWL/How-widespread-is-WhatsApps-usage-in-India.html}
\showURL{%
\tempurl}


\bibitem[\protect\citeauthoryear{Malliaros, Rossi, and Vazirgiannis}{Malliaros
  et~al\mbox{.}}{2016}]%
        {malliaros2016locating}
\bibfield{author}{\bibinfo{person}{Fragkiskos~D Malliaros},
  \bibinfo{person}{Maria-Evgenia~G Rossi}, {and} \bibinfo{person}{Michalis
  Vazirgiannis}.} \bibinfo{year}{2016}\natexlab{}.
\newblock \showarticletitle{Locating influential nodes in complex networks}.
\newblock \bibinfo{journal}{\emph{Scientific reports}}  \bibinfo{volume}{6}
  (\bibinfo{year}{2016}), \bibinfo{pages}{19307}.
\newblock


\bibitem[\protect\citeauthoryear{Mann and Whitney}{Mann and Whitney}{1947}]%
        {mann1947test}
\bibfield{author}{\bibinfo{person}{Henry~B Mann} {and}
  \bibinfo{person}{Donald~R Whitney}.} \bibinfo{year}{1947}\natexlab{}.
\newblock \showarticletitle{On a test of whether one of two random variables is
  stochastically larger than the other}.
\newblock \bibinfo{journal}{\emph{The annals of mathematical statistics}}
  (\bibinfo{year}{1947}), \bibinfo{pages}{50--60}.
\newblock


\bibitem[\protect\citeauthoryear{Mantel et~al\mbox{.}}{Mantel
  et~al\mbox{.}}{1966}]%
        {mantel1966evaluation}
\bibfield{author}{\bibinfo{person}{Nathan Mantel} {et~al\mbox{.}}}
  \bibinfo{year}{1966}\natexlab{}.
\newblock \showarticletitle{Evaluation of survival data and two new rank order
  statistics arising in its consideration}.
\newblock \bibinfo{journal}{\emph{Cancer Chemother Rep}} \bibinfo{volume}{50},
  \bibinfo{number}{3} (\bibinfo{year}{1966}), \bibinfo{pages}{163--170}.
\newblock


\bibitem[\protect\citeauthoryear{Mathew, Illendula, Saha, Sarkar, Goyal, and
  Mukherjee}{Mathew et~al\mbox{.}}{2020}]%
        {mathew2019temporal}
\bibfield{author}{\bibinfo{person}{Binny Mathew}, \bibinfo{person}{Anurag
  Illendula}, \bibinfo{person}{Punyajoy Saha}, \bibinfo{person}{Soumya Sarkar},
  \bibinfo{person}{Pawan Goyal}, {and} \bibinfo{person}{Animesh Mukherjee}.}
  \bibinfo{year}{2020}\natexlab{}.
\newblock \showarticletitle{Hate Begets Hate: A Temporal Study of Hate Speech}.
\newblock \bibinfo{journal}{\emph{Proc. ACM Hum.-Comput. Interact.}}
  \bibinfo{volume}{4}, \bibinfo{number}{CSCW2}, Article \bibinfo{articleno}{92}
  (\bibinfo{date}{Oct.} \bibinfo{year}{2020}), \bibinfo{numpages}{24}~pages.
\newblock


\bibitem[\protect\citeauthoryear{McLaughlin}{McLaughlin}{2018}]%
        {HowWhats65:online}
\bibfield{author}{\bibinfo{person}{Timothy McLaughlin}.}
  \bibinfo{year}{2018}\natexlab{}.
\newblock \bibinfo{title}{How WhatsApp Fuels Fake News and Violence in India |
  WIRED}.
\newblock
  \bibinfo{howpublished}{\url{https://www.wired.com/story/how-whatsapp-fuels-fake-news-and-violence-in-india/}}.
\newblock
\newblock
\shownote{(Accessed on 01/22/2021).}


\bibitem[\protect\citeauthoryear{Ousidhoum, Lin, Zhang, Song, and
  Yeung}{Ousidhoum et~al\mbox{.}}{2019}]%
        {ousidhoum2019multilingual}
\bibfield{author}{\bibinfo{person}{Nedjma Ousidhoum}, \bibinfo{person}{Zizheng
  Lin}, \bibinfo{person}{Hongming Zhang}, \bibinfo{person}{Yangqiu Song}, {and}
  \bibinfo{person}{Dit-Yan Yeung}.} \bibinfo{year}{2019}\natexlab{}.
\newblock \showarticletitle{Multilingual and Multi-Aspect Hate Speech
  Analysis}. In \bibinfo{booktitle}{\emph{Proceedings of the 2019 Conference on
  Empirical Methods in Natural Language Processing and the 9th International
  Joint Conference on Natural Language Processing (EMNLP-IJCNLP)}}.
  \bibinfo{pages}{4667--4676}.
\newblock


\bibitem[\protect\citeauthoryear{Parakh}{Parakh}{2017}]%
        {84DeadIn79:online}
\bibfield{author}{\bibinfo{person}{Rohit Parakh}.}
  \bibinfo{year}{2017}\natexlab{}.
\newblock \bibinfo{title}{84\% Dead In Cow-Related Violence Since 2010 Are
  Muslim; 97\% Attacks After 2014}.
\newblock
  \bibinfo{howpublished}{\url{https://www.indiaspend.com/86-dead-in-cow-related-violence-since-2010-are-muslim-97-attacks-after-2014-2014}}.
\newblock
\newblock
\shownote{(Accessed on 01/22/2021).}


\bibitem[\protect\citeauthoryear{Pereira-Kohatsu, Quijano-Sánchez, Liberatore,
  and Camacho-Collados}{Pereira-Kohatsu et~al\mbox{.}}{2019}]%
        {PMID:31717760}
\bibfield{author}{\bibinfo{person}{Juan~Carlos Pereira-Kohatsu},
  \bibinfo{person}{Lara Quijano-Sánchez}, \bibinfo{person}{Federico
  Liberatore}, {and} \bibinfo{person}{Miguel Camacho-Collados}.}
  \bibinfo{year}{2019}\natexlab{}.
\newblock \showarticletitle{Detecting and Monitoring Hate Speech in Twitter}.
\newblock \bibinfo{journal}{\emph{Sensors (Basel, Switzerland)}}
  \bibinfo{volume}{19}, \bibinfo{number}{21} (\bibinfo{date}{October}
  \bibinfo{year}{2019}).
\newblock
\urldef\tempurl%
\url{https://doi.org/10.3390/s19214654}
\showDOI{\tempurl}


\bibitem[\protect\citeauthoryear{Reis, Melo, Garimella, and Benevenuto}{Reis
  et~al\mbox{.}}{2020}]%
        {reis2020detecting}
\bibfield{author}{\bibinfo{person}{Julio~CS Reis}, \bibinfo{person}{Philipe
  Melo}, \bibinfo{person}{Kiran Garimella}, {and}
  \bibinfo{person}{Fabr{\'\i}cio Benevenuto}.} \bibinfo{year}{2020}\natexlab{}.
\newblock \showarticletitle{Can WhatsApp benefit from debunked fact-checked
  stories to reduce misinformation?}
\newblock \bibinfo{journal}{\emph{Harvard Kennedy School Misinformation
  Review}} (\bibinfo{year}{2020}).
\newblock


\bibitem[\protect\citeauthoryear{Resende, Melo, Sousa, Messias, Vasconcelos,
  Almeida, and Benevenuto}{Resende et~al\mbox{.}}{2019}]%
        {resende2019mis}
\bibfield{author}{\bibinfo{person}{Gustavo Resende}, \bibinfo{person}{Philipe
  Melo}, \bibinfo{person}{Hugo Sousa}, \bibinfo{person}{Johnnatan Messias},
  \bibinfo{person}{Marisa Vasconcelos}, \bibinfo{person}{Jussara Almeida},
  {and} \bibinfo{person}{Fabr{\'\i}cio Benevenuto}.}
  \bibinfo{year}{2019}\natexlab{}.
\newblock \showarticletitle{(Mis) Information Dissemination in WhatsApp:
  Gathering, Analyzing and Countermeasures}. In \bibinfo{booktitle}{\emph{The
  World Wide Web Conference}}. \bibinfo{pages}{818--828}.
\newblock


\bibitem[\protect\citeauthoryear{Ribeiro, Singh, and Guestrin}{Ribeiro
  et~al\mbox{.}}{2016}]%
        {ribeiro2016should}
\bibfield{author}{\bibinfo{person}{Marco~Tulio Ribeiro},
  \bibinfo{person}{Sameer Singh}, {and} \bibinfo{person}{Carlos Guestrin}.}
  \bibinfo{year}{2016}\natexlab{}.
\newblock \showarticletitle{" Why should I trust you?" Explaining the
  predictions of any classifier}. In \bibinfo{booktitle}{\emph{Proceedings of
  the 22nd ACM SIGKDD international conference on knowledge discovery and data
  mining}}.
\newblock


\bibitem[\protect\citeauthoryear{R\"{o}der, Both, and Hinneburg}{R\"{o}der
  et~al\mbox{.}}{2015}]%
        {10.1145/2684822.2685324}
\bibfield{author}{\bibinfo{person}{Michael R\"{o}der}, \bibinfo{person}{Andreas
  Both}, {and} \bibinfo{person}{Alexander Hinneburg}.}
  \bibinfo{year}{2015}\natexlab{}.
\newblock \showarticletitle{Exploring the Space of Topic Coherence Measures}.
  In \bibinfo{booktitle}{\emph{Proceedings of the Eighth ACM International
  Conference on Web Search and Data Mining}} (Shanghai, China)
  \emph{(\bibinfo{series}{WSDM '15})}. \bibinfo{publisher}{Association for
  Computing Machinery}, \bibinfo{address}{New York, NY, USA},
  \bibinfo{pages}{399–408}.
\newblock
\showISBNx{9781450333177}
\urldef\tempurl%
\url{https://doi.org/10.1145/2684822.2685324}
\showDOI{\tempurl}


\bibitem[\protect\citeauthoryear{Rosenbaum and Rubin}{Rosenbaum and
  Rubin}{1983}]%
        {10.1093/biomet/70.1.41}
\bibfield{author}{\bibinfo{person}{Paul~R. Rosenbaum} {and}
  \bibinfo{person}{Donald~B. Rubin}.} \bibinfo{year}{1983}\natexlab{}.
\newblock \showarticletitle{{The central role of the propensity score in
  observational studies for causal effects}}.
\newblock \bibinfo{journal}{\emph{Biometrika}} \bibinfo{volume}{70},
  \bibinfo{number}{1} (\bibinfo{date}{04} \bibinfo{year}{1983}),
  \bibinfo{pages}{41--55}.
\newblock
\showISSN{0006-3444}
\urldef\tempurl%
\url{https://doi.org/10.1093/biomet/70.1.41}
\showDOI{\tempurl}
\showeprint{https://academic.oup.com/biomet/article-pdf/70/1/41/662954/70-1-41.pdf}


\bibitem[\protect\citeauthoryear{Ross, Rist, Carbonell, Cabrera, Kurowsky, and
  Wojatzki}{Ross et~al\mbox{.}}{2017}]%
        {ross2017measuring}
\bibfield{author}{\bibinfo{person}{Bj{\"o}rn Ross}, \bibinfo{person}{Michael
  Rist}, \bibinfo{person}{Guillermo Carbonell}, \bibinfo{person}{Benjamin
  Cabrera}, \bibinfo{person}{Nils Kurowsky}, {and} \bibinfo{person}{Michael
  Wojatzki}.} \bibinfo{year}{2017}\natexlab{}.
\newblock \showarticletitle{Measuring the reliability of hate speech
  annotations: The case of the european refugee crisis}.
\newblock \bibinfo{journal}{\emph{Bochumer Linguistische Arbeitsberichte}}
  (\bibinfo{year}{2017}).
\newblock


\bibitem[\protect\citeauthoryear{Ruder, S{\o}gaard, and Vuli{\'c}}{Ruder
  et~al\mbox{.}}{2019}]%
        {ruder2019unsupervised}
\bibfield{author}{\bibinfo{person}{Sebastian Ruder}, \bibinfo{person}{Anders
  S{\o}gaard}, {and} \bibinfo{person}{Ivan Vuli{\'c}}.}
  \bibinfo{year}{2019}\natexlab{}.
\newblock \showarticletitle{Unsupervised cross-lingual representation
  learning}. In \bibinfo{booktitle}{\emph{Proceedings of the 57th Annual
  Meeting of the Association for Computational Linguistics: Tutorial
  Abstracts}}. \bibinfo{pages}{31--38}.
\newblock


\bibitem[\protect\citeauthoryear{Salminen, Hopf, Chowdhury, Jung, Almerekhi,
  and Jansen}{Salminen et~al\mbox{.}}{2020}]%
        {salminen2020developing}
\bibfield{author}{\bibinfo{person}{Joni Salminen}, \bibinfo{person}{Maximilian
  Hopf}, \bibinfo{person}{Shammur~A Chowdhury}, \bibinfo{person}{Soon-gyo
  Jung}, \bibinfo{person}{Hind Almerekhi}, {and} \bibinfo{person}{Bernard~J
  Jansen}.} \bibinfo{year}{2020}\natexlab{}.
\newblock \showarticletitle{Developing an online hate classifier for multiple
  social media platforms}.
\newblock \bibinfo{journal}{\emph{Human-centric Computing and Information
  Sciences}} \bibinfo{volume}{10}, \bibinfo{number}{1} (\bibinfo{year}{2020}),
  \bibinfo{pages}{1}.
\newblock


\bibitem[\protect\citeauthoryear{Sampietro}{Sampietro}{2019}]%
        {sampietro2019emoji}
\bibfield{author}{\bibinfo{person}{Agnese Sampietro}.}
  \bibinfo{year}{2019}\natexlab{}.
\newblock \showarticletitle{Emoji and rapport management in Spanish WhatsApp
  chats}.
\newblock \bibinfo{journal}{\emph{Journal of Pragmatics}}
  \bibinfo{volume}{143} (\bibinfo{year}{2019}), \bibinfo{pages}{109--120}.
\newblock


\bibitem[\protect\citeauthoryear{Sarkar and Sarkar}{Sarkar and Sarkar}{2016}]%
        {sarkar2016sacred}
\bibfield{author}{\bibinfo{person}{Radha Sarkar} {and} \bibinfo{person}{Amar
  Sarkar}.} \bibinfo{year}{2016}\natexlab{}.
\newblock \showarticletitle{Sacred slaughter: An analysis of historical,
  communal, and constitutional aspects of beef bans in India}.
\newblock \bibinfo{journal}{\emph{Politics, Religion \& Ideology}}
  \bibinfo{volume}{17}, \bibinfo{number}{4} (\bibinfo{year}{2016}),
  \bibinfo{pages}{329--351}.
\newblock


\bibitem[\protect\citeauthoryear{Seidler, Vondr{\'a}{\v{c}}ek, and
  Saxl}{Seidler et~al\mbox{.}}{2000}]%
        {seidler2000life}
\bibfield{author}{\bibinfo{person}{Jan Seidler},
  \bibinfo{person}{Ji{\v{r}}{\'\i} Vondr{\'a}{\v{c}}ek}, {and}
  \bibinfo{person}{Ivan Saxl}.} \bibinfo{year}{2000}\natexlab{}.
\newblock \showarticletitle{The life and work of Zbyn{\v{e}}k {\v{S}}id{\'a}k
  (1933--1999)}.
\newblock \bibinfo{journal}{\emph{Applications of Mathematics}}
  \bibinfo{volume}{45}, \bibinfo{number}{5} (\bibinfo{year}{2000}),
  \bibinfo{pages}{321--336}.
\newblock


\bibitem[\protect\citeauthoryear{Shin, Eliassi-Rad, and Faloutsos}{Shin
  et~al\mbox{.}}{2016}]%
        {shin2016corescope}
\bibfield{author}{\bibinfo{person}{Kijung Shin}, \bibinfo{person}{Tina
  Eliassi-Rad}, {and} \bibinfo{person}{Christos Faloutsos}.}
  \bibinfo{year}{2016}\natexlab{}.
\newblock \showarticletitle{Corescope: Graph mining using k-core
  analysis—patterns, anomalies and algorithms}. In
  \bibinfo{booktitle}{\emph{2016 IEEE 16th International Conference on Data
  Mining (ICDM)}}. IEEE, \bibinfo{pages}{469--478}.
\newblock


\bibitem[\protect\citeauthoryear{Sindoni}{Sindoni}{2018}]%
        {fear_work}
\bibfield{author}{\bibinfo{person}{Maria~Grazia Sindoni}.}
  \bibinfo{year}{2018}\natexlab{}.
\newblock \showarticletitle{Direct hate speech vs. indirect fear speech. A
  multimodal critical discourse analysis of the Sun's editorial "1 in 5 Brit
  Muslims' sympathy for jihadis"}.
\newblock \bibinfo{journal}{\emph{Lingue e Linguaggio}}  \bibinfo{volume}{28}
  (\bibinfo{date}{12} \bibinfo{year}{2018}), \bibinfo{pages}{267--292}.
\newblock


\bibitem[\protect\citeauthoryear{Tekiro{\u{g}}lu, Chung, and
  Guerini}{Tekiro{\u{g}}lu et~al\mbox{.}}{2020}]%
        {sinem2020generating}
\bibfield{author}{\bibinfo{person}{Serra~Sinem Tekiro{\u{g}}lu},
  \bibinfo{person}{Yi-Ling Chung}, {and} \bibinfo{person}{Marco Guerini}.}
  \bibinfo{year}{2020}\natexlab{}.
\newblock \showarticletitle{Generating Counter Narratives against Online Hate
  Speech: Data and Strategies}. In \bibinfo{booktitle}{\emph{Proceedings of the
  58th Annual Meeting of the Association for Computational Linguistics}}.
  \bibinfo{publisher}{Association for Computational Linguistics},
  \bibinfo{pages}{1177--1190}.
\newblock


\bibitem[\protect\citeauthoryear{Wilkinson, Dumontier, Aalbersberg, Appleton,
  Axton, Baak, Blomberg, Boiten, da~Silva~Santos, Bourne,
  et~al\mbox{.}}{Wilkinson et~al\mbox{.}}{2016}]%
        {wilkinson2016fair}
\bibfield{author}{\bibinfo{person}{Mark~D Wilkinson}, \bibinfo{person}{Michel
  Dumontier}, \bibinfo{person}{IJsbrand~Jan Aalbersberg},
  \bibinfo{person}{Gabrielle Appleton}, \bibinfo{person}{Myles Axton},
  \bibinfo{person}{Arie Baak}, \bibinfo{person}{Niklas Blomberg},
  \bibinfo{person}{Jan-Willem Boiten}, \bibinfo{person}{Luiz~Bonino da
  Silva~Santos}, \bibinfo{person}{Philip~E Bourne}, {et~al\mbox{.}}}
  \bibinfo{year}{2016}\natexlab{}.
\newblock \showarticletitle{The FAIR Guiding Principles for scientific data
  management and stewardship}.
\newblock \bibinfo{journal}{\emph{Scientific data}} \bibinfo{volume}{3},
  \bibinfo{number}{1} (\bibinfo{year}{2016}), \bibinfo{pages}{1--9}.
\newblock


\bibitem[\protect\citeauthoryear{Workneh}{Workneh}{2019}]%
        {doi:10.1080/23743670.2020.1729832}
\bibfield{author}{\bibinfo{person}{Téwodros~W. Workneh}.}
  \bibinfo{year}{2019}\natexlab{}.
\newblock \showarticletitle{Ethiopia’s Hate Speech Predicament: Seeking
  Antidotes Beyond a Legislative Response}.
\newblock \bibinfo{journal}{\emph{African Journalism Studies}}
  \bibinfo{volume}{40}, \bibinfo{number}{3} (\bibinfo{year}{2019}).
\newblock


\bibitem[\protect\citeauthoryear{Yasseri and Vidgen}{Yasseri and
  Vidgen}{2019}]%
        {yasseri2019detecting}
\bibfield{author}{\bibinfo{person}{T Yasseri} {and} \bibinfo{person}{B
  Vidgen}.} \bibinfo{year}{2019}\natexlab{}.
\newblock \showarticletitle{Detecting weak and strong Islamophobic hate speech
  on social media}.
\newblock \bibinfo{journal}{\emph{Journal of Information Technology and
  Politics}} \bibinfo{volume}{17}, \bibinfo{number}{1} (\bibinfo{year}{2019}).
\newblock


\bibitem[\protect\citeauthoryear{Ybarra, Mitchell, Wolak, and Finkelhor}{Ybarra
  et~al\mbox{.}}{2006}]%
        {ybarra2006examining}
\bibfield{author}{\bibinfo{person}{Michele~L Ybarra},
  \bibinfo{person}{Kimberly~J Mitchell}, \bibinfo{person}{Janis Wolak}, {and}
  \bibinfo{person}{David Finkelhor}.} \bibinfo{year}{2006}\natexlab{}.
\newblock \showarticletitle{Examining characteristics and associated distress
  related to Internet harassment: findings from the Second Youth Internet
  Safety Survey}.
\newblock \bibinfo{journal}{\emph{Pediatrics}} \bibinfo{volume}{118},
  \bibinfo{number}{4} (\bibinfo{year}{2006}).
\newblock


\bibitem[\protect\citeauthoryear{Zhang, Robinson, and Tepper}{Zhang
  et~al\mbox{.}}{2018}]%
        {zhang2018detecting}
\bibfield{author}{\bibinfo{person}{Ziqi Zhang}, \bibinfo{person}{David
  Robinson}, {and} \bibinfo{person}{Jonathan Tepper}.}
  \bibinfo{year}{2018}\natexlab{}.
\newblock \showarticletitle{Detecting hate speech on twitter using a
  convolution-gru based deep neural network}. In
  \bibinfo{booktitle}{\emph{European semantic web conference}}. Springer,
  \bibinfo{pages}{745--760}.
\newblock


\end{thebibliography}

\end{document}